\documentclass[]{aa}  

\usepackage{graphicx}
\usepackage{txfonts}
\usepackage{natbib} 
\bibpunct{(}{)}{;}{a}{}{,} 

\usepackage{verbatim}
\usepackage{url}
\usepackage{epstopdf}

\begin{document}

   \title{Statistical errors and systematic biases in the calibration of the convective core overshooting with eclipsing binaries} 

   \subtitle{A case study: TZ~Fornacis}

   \author{G. Valle \inst{1,2,3}, M. Dell'Omodarme \inst{3}, P.G. Prada Moroni
     \inst{2,3}, S. Degl'Innocenti \inst{2,3} 
          }
   \titlerunning{Calibration of core overshooting with TZ~Fornacis}
   \authorrunning{Valle, G. et al.}

   \institute{
INAF - Osservatorio Astronomico di Collurania, Via Maggini, I-64100, Teramo, Italy 
\and
 INFN,
 Sezione di Pisa, Largo Pontecorvo 3, I-56127, Pisa, Italy
\and
Dipartimento di Fisica "Enrico Fermi'',
Universit\`a di Pisa, Largo Pontecorvo 3, I-56127, Pisa, Italy
 }

   \offprints{G. Valle, valle@df.unipi.it}

   \date{Received 3 February 2016; accepted 13 December 2016}

  \abstract
{Recently published work has made high-precision fundamental parameters available for the binary system TZ~Fornacis, making it an ideal target for the calibration of stellar models.}
   {Relying on these observations, we attempt to constrain the initial helium abundance, the age and the efficiency of the convective core overshooting. Our main aim is in pointing out the  biases in the results due to not accounting for some sources of uncertainty. }
{       
We adopt the SCEPtER pipeline, a maximum likelihood technique  based on fine grids of stellar models computed for various values of metallicity, initial helium abundance and overshooting efficiency by means 
 of two independent stellar evolutionary codes, namely FRANEC and MESA. 
}
  {Beside the degeneracy between the estimated age and overshooting efficiency, we found the existence of multiple independent groups of solutions. The best one suggests a system of age 1.10 $\pm$ 0.07 Gyr 
  composed of a primary star in the central helium burning stage and a secondary in the sub-giant branch (SGB). The resulting initial helium abundance is consistent with a helium-to-metal enrichment ratio 
  of  $\Delta Y/\Delta Z = 1$; the core overshooting parameter is $\beta = 0.15 \pm 0.01$ for FRANEC and $f_{\rm ov}= 0.013 \pm 0.001 $ for MESA. The second class of solutions, characterised by a worse goodness-of-fit,  still 
  suggest a primary star in the central helium-burning stage but a secondary in the overall contraction phase, at the end of the main sequence (MS). In this case, the FRANEC grid provides an age of $1.16_{-0.02}^{+0.03}$ Gyr 
  and a core overshooting parameter $\beta = 0.25_{-0.01}^{+0.005}$, while the MESA grid gives $1.23 \pm 0.03$ Gyr and $f_{\rm ov} = 0.025 \pm 0.003$. 
  We analyse the impact on the results of a larger, but typical, mass uncertainty and of neglecting the uncertainty in the initial helium content of the system.
  We show that very precise mass determinations with uncertainty  of a few thousandths of solar mass are required to obtain reliable determinations of stellar parameters, as 
  mass errors larger than approximately 1\% lead to estimates that are not only less precise but also biased. Moreover, we show that a fit obtained with a grid of models computed at a fixed $\Delta Y/\Delta Z $ --  thus neglecting 
  the current uncertainty in the initial helium content of the system -- can provide severely biased age and overshooting estimates. The possibility of independent overshooting efficiencies for the two stars of the system  is also explored.
 }
{The present analysis confirms that to constrain the core overshooting parameter by means of binary systems 
is a very difficult task that requires an observational precision still rarely achieved and a robust statistical treatment of the error sources.}


   \maketitle

\section{Introduction}\label{sec:intro}

Stellar models play a central role in many astronomical research fields as they are
 routinely used to infer important physical quantities. In spite of their significant improvement 
 over the last decades, stellar models are still affected by non-negligible uncertainties. 
One of the major and long standing weaknesses is the lack of a rigorous treatment of
convective transport \citep[see][for a comprehensive introduction]{Viallet2015}.  As a consequence, both the extension of convective regions and the 
temperature profile inside them is not yet a robust prediction of the current generation of stellar models.  Within such a scenario, to compute 
the dimension of the convective core, one usually determines the classical Schwarzschild border and then 
allows for an overshooting region whose extension is a function of a free parameter. It is thus clear that the convective core mass (hereafter we refer to the fully mixed central region as the convective core), 
which has a profound impact on many evolutionary features, can not be firmly predicted. On the other hand, 
the overshooting extension can be empirically calibrated. 

In this regard, the importance of double-lined eclipsing binary systems for stellar evolution models is well recognised in the literature \citep[see, among many,][]{Andersen1991,Torres2010}. However, even after the great improvement in observations, which nowadays routinely reach precisions
 in stellar mass and radius determinations of a few percent, the error in the estimated age of the system is seldom lower than 10-15\%, mainly owing to poorly constrained parameters such as the  initial  helium abundance and the efficiency of the convective core overshooting in the main-sequence (MS) phase \citep{binary}. Recent studies, focused on the calibration of the overshooting extension in binary systems \citep{Claret2007,Stancliffe2015}, show that a large uncertainty is still present.  
 Moreover, in a recent theoretical investigation, we showed that the overshooting calibration based on eclipsing binaries with both members in the MS phase, with error in mass determination at the percent level, are affected by statistical errors and systematic biases large enough to hamper a statistically meaningful conclusion, thus 
 suggesting great caution must be taken \citep{overshooting}. It is apparent that only very precise observations of binary systems could, in principle, significantly improve model calibration.  

A favourable opportunity was recently provided by \citet{Gallenne2015}, who made available estimates of the stellar masses in the double-line eclipsing binary system TZ~Fornacis with the unprecedented precision of 0.001 $M_{\sun}$. \citet{Gallenne2015} adopted their determination to estimate the age of the system for two assumed values of the initial metallicity.

A binary system such as this, where the secondary star is still in the sub-giant branch (SGB) or earlier while the primary is already in a more advanced stage, offers the opportunity 
to constrain some other parameters besides the age, such as the initial helium abundance and the convective core overshooting extension, much more efficiently than 
in binary systems where both components are in MS. 
On the theoretical side, a twofold heavy computational 
effort is necessary to provide, firstly, a large and fine grid of stellar models spanning a range of plausible values of the parameters to be constrained and, secondly, a robust 
statistical procedure for evaluating the statistical error affecting the inferred parameters.

For our purposes, the main interest of this system is that it provides a case study for analysing the difficulty in obtaining a sensible calibration of unconstrained stellar parameters, even in the presence of very precise observational data. In this respect, we aim to show how the different uncertainty sources in stellar models propagate into the final calibration and how multiple detached solutions can exist in the hyperspace of the explored stellar parameters. We are also interested in analysing the biases in the overshooting calibration that can be induced by the adoption of a grid of stellar models that arbitrarily neglects the variability of some model parameters, such as the initial helium abundance.
Moreover, the TZ~Fornacis data can be exploited to test stellar evolution models. In particular, due to the negligible uncertainty on mass determination, it is realistic to test several assumptions adopted in the binary system modelling. In this work we assume, as is usual, that the two stars in the detached system can be modelled as two independent objects, therefore neglecting the tidal interaction between them. 
This hypothesis, in addition to the fact that the two stars have almost equal mass (mass ratio 1.05), allows us to suppose that the stars share a common overshooting efficiency. An impossibility to obtain a fit in this scenario will point out some difficulty in the adopted stellar models and/or in the stated assumptions.

In the following discussion we do not analyse the impact of the current uncertainty affecting the input physics required
to compute stellar models (i.e. EOS, radiative and conductive opacities,
nuclear reaction cross sections, etc.) on overshooting calibration. In fact we kept fixed the input physics to their reference values. Therefore, the reported uncertainties in the calibrated parameters (age, initial helium abundance and overshooting efficiency) come only from the propagation of the observational errors in the observables used to constrain the fitting procedure ($M$, $R$, $T_{\rm eff}$, [Fe/H]).

The structure of the paper is as follows. In Sect.~\ref{sec:method}, we 
discuss the method and the grids used in the estimation process. 
The best fit for the system is
 presented in Sect.~\ref{sec:results}. 
The effect of varying the observational errors is explored in Sect.~\ref{sec:errori-obs}. The impact of changing the prior constraints in the maximum likelihood procedure is discussed in Sect.~\ref{sec:ML-priors}. 
 The results of the analysis are compared with the literature in Sect.~\ref{sec:literature}.
Some concluding remarks can be found in Sect.~\ref{sec:conclusions}.

\section{Methods}\label{sec:method}

The estimation process is based on a modified SCEPtER pipeline\footnote{Publicly available on CRAN: \url{http://CRAN.R-project.org/package=SCEPtER}, \url{http://CRAN.R-project.org/package=SCEPtERbinary}}, a well as a tested grid-based maximum likelihood technique, which has been adopted in the past for single stars \citep{scepter1,eta,bulge} and binary systems \citep{binary}. Briefly, the procedure computes a likelihood value for all the grid points and it provides estimates of the parameters of interest (age, initial helium abundance, initial metallicity, core overshooting parameter  and extension of the convective core) by averaging the values of all the models with likelihood greater than 95\% of the maximum value.

We assume
${\cal S}_1$ and ${\cal S}_2$ to be two stars in a detached binary system. The observed quantities adopted as observational constraints are $q^{{\cal S}_{1,2}} \equiv \{T_{\rm eff, {\cal S}_{1,2}}, {\rm
        [Fe/H]}_{{\cal S}_{1,2}}, 
M_{{\cal S}_{1,2}}, R_{{\cal S}_{1,2}}\}$. Let $\sigma^{1,2} = \{\sigma(T_{\rm
        eff, {\cal S}_{1,2}}), \sigma({\rm [Fe/H]}_{{\cal S}_{1,2}}), \sigma(M_{{\cal S}_{1,2}}),
\sigma(R_{{\cal S}_{1,2}})\}$ be the uncertainty in the observed
quantities. For each point $j$ on the estimation grid of stellar models, 
we define $q^{j} \equiv \{T_{{\rm eff}, j}, {\rm [Fe/H]}_{j}, M_{j},
R_{j}\}$. 
Let $ {{\cal L}^{1,2}}_j$ be the single-star likelihood functions defined as
\begin{equation}
        {{\cal L}^{1,2}}_j = \left( \prod_{i=1}^4 \frac{1}{\sqrt{2 \pi}
                \sigma^{1,2}_i} \right) 
        \times \exp \left( -\frac{\chi_{1,2}^2}{2} \right)
        \label{eq:lik}
        ,\end{equation}
where
\begin{equation}
        \chi^2_{1,2} = \sum_{i=1}^4 \left( \frac{q_i^{{\cal S}_{1,2}} -
                q_i^j}{\sigma_i} \right)^2. 
        \label{eq:chi2}
\end{equation}

Then, the joint likelihood  ${\cal \tilde L}$ of the system is computed as the product of the single star likelihood functions.
In the computation of the likelihood functions, we consider only the grid points  
within $3 \sigma$ of all the variables from the observational constraints.
The pipeline provides the estimates both for the individual components and for 
the  whole system. In the former case, the fits for the two stars are obtained independently,
while in the latter case the algorithm imposes that the two members must have a common age (with a tolerance of 1 Myr), identical initial helium abundance and initial metallicity and a  common overshooting efficiency parameter. 
We discuss this last assumption in more detail in Sect.~\ref{sec:diff-ov}.

The estimation process relies on stellar models spanning a wide range of evolutionary phases with very different time scales; therefore the time step between consecutive points is far from uniform. As a consequence, grid based estimates can be biased towards more densely represented  phases. To obtain a sensible estimate of the density function of the age of the system and of the best overshooting parameter consistent with the data, we adopt a weighted approach \citep[see e.g.][]{Jorgensen2005, eta}. Each point in the grid is weighted with the evolutionary time step around it and this weight is inserted as a multiplicative factor in Eq.~(\ref{eq:lik}). Such a procedure intrinsically favours lower helium abundance and higher metallicity models, because their evolution is slower. A different weighting scheme was also tried, allowing for a normalization at track level so that the highest weight for each track is equal to one, with negligible differences in the results.

\begin{table}[ht]
        \centering
        \caption{Observational constraints for the TZ~Fornacis binary system from \citet{Gallenne2015}, but with stellar radii from \citet{Andersen1991}.}
        \label{tab:input}
        \begin{tabular}{lcc}
                \hline\hline
                & primary & secondary \\
                \hline 
                $M$ ($M_{\sun}$) & 2.057 $\pm$ 0.001 & 1.958 $\pm$ 0.001\\
                $R$ ($R_{\sun}$) & 8.32 $\pm$ 0.12 & 3.96 $\pm$ 0.09\\
                $T_{\rm eff}$ (K)  & 4930 $\pm$ 30 & 6650 $\pm$ 200\\
                ${\rm [Fe/H]}$ & 0.02 $\pm$ 0.05 & $-0.05$ $\pm$ 0.1\\
                \hline
        \end{tabular}
\end{table}

As observational constraints,  we use the masses, radii, metallicities [Fe/H] and effective temperatures of both stars. The adopted values and their uncertainties, reported in Table~\ref{tab:input}, are taken from \citet{Gallenne2015} except for the radii, for which we adopt the more precise photometric estimate from \citet{Andersen1991}.
Apart from the high accuracy in mass determination,  \citet{Gallenne2015} also provides a precise estimate of the effective temperature of the primary star, with a nominal uncertainty of 30 K. This is the value we adopted in our analysis. However, in order to take into account possible systematic uncertainties on the effective temperature scale determination, we also performed a more conservative analysis assuming an uncertainty of 100 K,  showing only minor differences. 

The error on the estimated parameters is obtained by means of Monte Carlo simulations. 
We generate $N = 10\,000$ artificial binary systems, sampling from a multivariate Gaussian distribution with mean $\mu = \{q^{S_1}, q^{S_2}\}$ and covariance matrix $\Sigma$.
Since the observationally inferred values
of a given physical quantity for the two binary components
are correlated, it would be unsafe to adopt a diagonal covariance matrix. Off-diagonal elements computed adopting sensible correlation coefficients should be included in
$\Sigma$ whenever a realistic noise needs to be simulated.
We assume a correlation of 0.95 between
the primary and secondary effective temperatures, and 0.95
between the metallicities of the two stars. Regarding mass correlation, the high precision of the estimate makes this parameter of no importance, but we set it at 0.8, a typical value for this class of stars \citep{binary}.  The correlation between the  radii is set at $-0.9$. Since the correlation between the observed radii was not provided in the analysis by \citet{Andersen1991}, we derived it by error propagation assuming normality, as in \citet{binary}. 

For each of these $N$ systems we repeat the parameter estimates. The errors on the parameters constrained by the fit (i.e. age, initial helium abundance, metallicity and overshooting)
are assessed by their multivariate posterior densities as detailed below.
As such, the obtained error accounts only for the propagation of the uncertainty in the observational constraints and neglects the systematic due to the stellar models. All the obtained solutions are validated by means of a $\chi^2$ test to assess their agreement with the observational data.

\subsection{Stellar model grids}
\label{sec:grids}

The grids of models were computed for the  masses of the two stars, that is,
$M_1 = 2.057$ $M_{\sun}$ and $M_2 = 1.958$ $M_{\sun}$  from the zero-age MS
up to the red-giant branch (RGB) ascension for the less massive star and to the
central helium depletion for the more massive component. 
The error on masses was neglected since a shift of 0.001 $M_{\sun}$ has negligible effect on the parameters of interest, 
but see Sect.~\ref{sec:errorM} for a discussion of the effects of mass error of approximately 0.01 $M_{\sun}$ or larger.
The initial metallicity [Fe/H] was varied from $-0.1$ dex to 0.1 dex, with
a step of 0.025 dex. 
The solar heavy-element mixture by \citet{AGSS09} was adopted. 
Several initial helium abundances were considered at fixed metallicity by adopting the commonly used
linear relation $Y = Y_p+\frac{\Delta Y}{\Delta Z} Z$
with the primordial abundance $Y_p = 0.2485$ from WMAP
\citep{peimbert07a,peimbert07b} and with a helium-to-metal enrichment ratio $\Delta Y/\Delta Z$
from 1 to 3 with a step of 0.5 \citep{gennaro10}. 

To check the robustness of the results and their dependence on the adopted stellar evolutionary code, we 
computed two grids of models by means of independent codes, namely FRANEC \citep{scilla2008, Tognelli2011} and MESA
 \citep[][release 7184]{MESA2013}.

The FRANEC code was used in the same
configuration as was adopted to compute the Pisa Stellar
Evolution Data Base\footnote{\url{http://astro.df.unipi.it/stellar-models/}} 
for low-mass stars \citep{database2012}. 
The models were computed
by assuming the solar-scaled mixing-length parameter $\alpha_{\rm
ml} = 1.74$. The calibration is performed repeating the Sun evolution by changing $Z$, $Y$ and $\alpha_{\rm ml}$. The iteration stops when, at the Sun age, the computed radius, luminosity, effective temperature, and photospheric [Fe/H] match the observed values with relative tolerance $10^{-4}$.
The extension of the extra-mixing region beyond the Schwarzschild border
was parametrized  in terms of the pressure scale height $H_{\rm 
        p}$: $l_{\rm ov} = \beta H_{\rm p}$, with 
$\beta$ in the range
[0.00; 0.30] with a step of 0.01. The code adopts an instantaneous mixing in the overshooting treatment. 
Semiconvection during the central He-burning phase 
\citep{castellani1971} was treated according to the algorithm
described in \citet{castellani1985}, resulting in no additional free parameters to be tuned. Breathing pulses were suppressed \citep{castellani1985, cassisi2001} as suggested by \citet{caputo1989}.  Atmospheric models by \citet{brott05}, computed using the PHOENIX code
        \citep{hauschildt99,hauschildt03},  
        available in the range $3000 \; {\rm K} \le T_{\rm eff} \le 10000 \;
        {\rm K}$, $0.0 \le \log g \; {\rm (cm \; s^{-2})} \le 5.0$, and $-4.0 \le {\rm
                [M/H]} 
        \le 0.5$ where adopted.
        In the range $10000 \; {\rm K} \le T_{\rm eff} \le 50000 \; {\rm K}$, $0.0
        \le \log g \; {\rm (cm \; s^{-2})} \le 5.0$, and $-2.5 \le {\rm [M/H]} 
        \le 0.5$, where models from \citet{brott05} are unavailable,    
        models by \citet{castelli03} are used.
Further details on the stellar models are fully
described in \citet{eta,binary} and references therein.  

The MESA computations assumed the same set of opacities and initial chemical compositions as in FRANEC ones. Outer boundary conditions are slightly different as MESA adopts the
atmospheric models from \citep{hauschildt99}, supplemented with models by \citet{castelli03} where the former are unavailable. The difference in the atmospheric models causes a small shift in the theoretical tracks, MESA models being approximately 20 K cooler than FRANEC ones in the RGB phase. In MESA, we obtained a value of the solar calibrated mixing-length  $\alpha_{\rm
ml} = 1.78$, in agreement with that reported by \citet{Stancliffe2016}. The code uses a different scheme for the convective core overshooting, relying on a diffusive approach with exponential decay of the overshooting diffusion coefficient.
In particular, MESA sets the overshoot mixing diffusion coefficient $D_{\rm OV}$ by:
\begin{equation}
D_{\rm OV} = D_{\rm conv} \exp{\left( -\frac{2 \; z}{f_{\rm ov} \; H_{\rm p}} \right)}
\label{eq:mesadiffusion}
,\end{equation}
where $D_{\rm conv}$ is the diffusion coefficient from mixing-length theory at a
location near the Schwarzschild boundary, $z$ is the distance in the
radiative layer from that location and $f_{\rm ov}$ is a free parameter. 
The overshooting parameter adopted in the diffusive scheme in MESA and the instantaneous one in FRANEC approximately follow the relation $f_{\rm ov} \approx \beta/10$ \citep{Noels2010}.
The different treatment of the overshooting adopted by FRANEC and MESA could shed light on possible systematic differences in the evaluation of the evolutionary time scale and on the convective core mass.  
We computed a grid that spans an overshooting efficiency in MS in the range $f_{\rm ov} \in$ [0.00; 0.04] with a step of 0.001.
Contrarily to FRANEC code, MESA also adopts a core overshooting scheme during the central helium burning phase. Since our aim is to calibrate the overshooting 
extension during the MS phase, we computed the models of the MESA grid by keeping the overshooting parameter fixed to $f_{\rm ov}$ = 0.014 during 
the central helium-burning phase. Although the effects of such a choice are expected to be very minor (see the discussion in Sect.~\ref{sec:results}),
 we also built a grid with a crude assumption of no overshooting in the central helium burning phase for a further consistency check. For all the input and configurations not explicitly mentioned above, we 
 adopted the default choices in standard MESA release.  
An analysis of the differences that can be expected for assuming different codes and input in stellar evolution computation for stars of 1 and 3 $M_{\sun}$ can be found in \citet{Stancliffe2016}.

The estimates of the stellar masses provided by \citet{Gallenne2015} are so precise (0.001 $M_{\sun}$) that their contribution to the statistical error in the  
stellar age, initial helium abundance and overshooting parameter can be safely neglected. However, much larger errors ($\gtrsim 0.01$ $M_{\sun}$) are still quite common in modern binary system observations. 
It is thus worth discussing the effect of mass errors of this order of magnitude on the parameter calibration. To this aim, we computed an additional grid of stellar models with the FRANEC 
code for 18 values of the mass in the range [1.88; 2.15] $M_{\sun}$, and the same chemical compositions (initial metallicity and helium abundance) and core overshooting parameters previously specified. 
This multi-mass grid of models was then adopted to recover the stellar parameters by means of the SCEPtER pipeline as described in Sect.~\ref{sec:method} but assuming larger 
uncertainties in the stellar masses. This exercise was performed twice. A first simulation adopted the uncertainties on the masses
 (1.5\% for the secondary and 3\% for the primary) from \citet{Andersen1991}, that is, the most precise values before the analysis by \citet{Gallenne2015}.
In a second simulation, a mass uncertainty of 1\% for both stars was adopted, a typical value that is nowadays routinely reached. 

A total of approximately 64\,000 stellar tracks were computed for this work.

\section{Parameters best estimates}
\label{sec:results}

In this section we try to calibrate the age, core overshooting efficiency and initial helium abundance, taking the observational errors at face value. The effect of changing their extent is discussed in Sect.~\ref{sec:errori-obs}.

Although the two evolutionary codes used to compute stellar models are independent, the results obtained by means of FRANEC and MESA grids are 
similar. Interestingly, both grids of stellar models provided multiple solutions in spite of the very precise mass estimates 
of the two stars. This result is partially due to the peculiar position of the secondary star, which can be fitted either in the MS overall contraction phase or in the SGB, 
as detailed below. Overall, it demonstrates, in practice, the high degree of precision in the observations that is needed to unambiguously constrain the free parameters governing stellar evolution. 

\begin{figure*}
        \centering
        \includegraphics[width=8.cm,angle=-90]{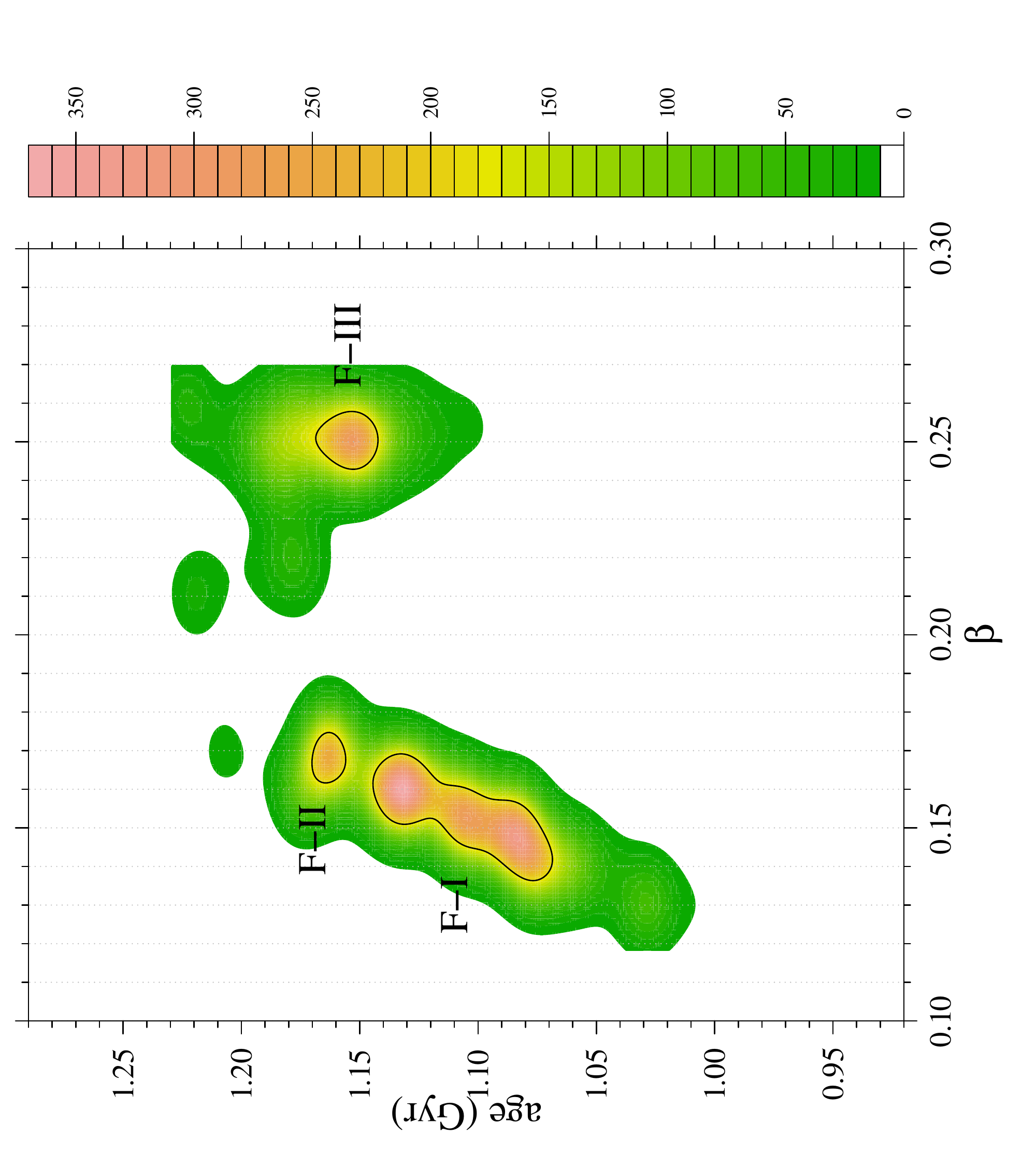} 
        \includegraphics[width=8.cm,angle=-90]{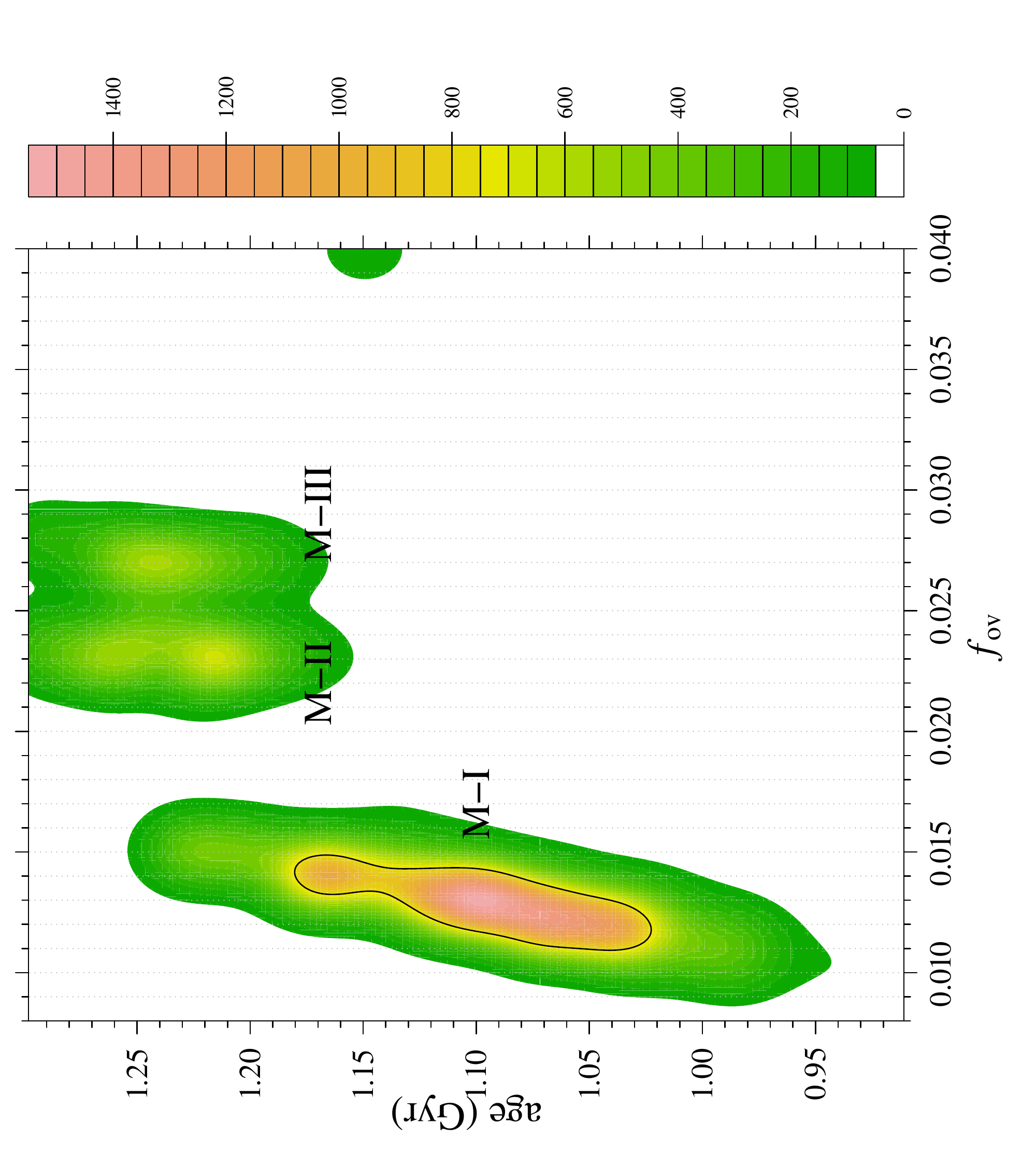} 
        \caption{{\it Left}: joint two-dimensional density of probability for the estimated overshooting parameter $\beta$ and the age of the binary system. The estimates were obtained by relying on FRANEC stellar models. The solid black line corresponds to points for which the density is half of the maximum value. {\it Right}: same as in the {\it left panel}, but for estimates from the MESA grid. }
        \label{fig:franec_mesa-ov-age}
\end{figure*}

Figure~\ref{fig:franec_mesa-ov-age} shows the two-dimensional density of probability of the estimated core overshooting efficiency and system age. 
As can easily be seen, in addition to the well-known degeneracy between age and overshooting, both FRANEC and MESA grids provide two detached islands of solutions. It is worth noting that the colour-scale in the two panels is different. In fact, all the densities are normalised to one; since different grids provide different relative height of the peak in the overshooting vs. age plane, the overall maximum is obviously different from one grid to another. Therefore, a comparison of the relative height is only meaningful within the individual panels and not between them.  

\begin{table*}[ht]
        \centering
        \caption{Multiple solutions for the  TZ~Fornacis binary system from the FRANEC grid of stellar models.}\label{tab:sol-FRANEC}
        \begin{tabular}{l|ccc|ccc|ccc|ccc}
                \hline\hline
                & \multicolumn{3}{c|}{F-I} & \multicolumn{3}{c|}{F-II} & \multicolumn{3}{c|}{F-I,II} & \multicolumn{3}{c}{F-III}\\
                &  $q_{16}$ & $q_{50}$ & $q_{84}$ &  $q_{16}$ & $q_{50}$ & $q_{84}$ &  $q_{16}$ & $q_{50}$ & $q_{84}$ &  $q_{16}$ & $q_{50}$ & $q_{84}$  \\ 
                \hline
                $Y$ & 0.261 & 0.262 & 0.263 & 0.263 & 0.264 & 0.264 &  0.261 & 0.262 & 0.263 & 0.262 & 0.263 & 0.264 \\ 
                $Z$ &  0.012 & 0.013 & 0.014 & 0.015 & 0.015 & 0.016 & 0.013 & 0.013 & 0.015  & 0.014 &   0.015 & 0.016 \\ 
                $\beta$ &  0.140 & 0.151 & 0.160 & 0.160 & 0.170 & 0.170 &  0.141 & 0.155 & 0.163 & 0.240 & 0.250 & 0.255 \\ 
                age (Gyr) & 1.07 & 1.10 & 1.13 & 1.16 & 1.16 & 1.17 &  1.08 & 1.11 &   1.16 &  1.14 & 1.16 & 1.19 \\ 
                $M_{\rm cc}$ ($M_{\sun}$) & - & - & - & - & - & - & - & - & - & 0.169 & 0.171 & 0.174 \\ 
                \hline
\multicolumn{13}{c}{Fit parameters}\\
\hline
$T_{\rm eff,1}$ (K) &  & 4952 &  &  & 4917 &  &  & 4947 &  &  & 4914 &  \\  
$T_{\rm eff,2}$ (K) &  & 6896$^*$ &  &  & 6788 &  &  & 6875$^*$ &  &  & 6701 &  \\
$R_1$ ($R_{\sun}$)&  & 8.39 &  &  & 8.43 &  &  & 8.39 &  &  & 8.61$^*$ &  \\ 
$R_2$ ($R_{\sun}$)&  & 4.03 &  &  & 4.04 &  &  & 4.03 &  &  & 3.71$^{**}$ &  \\
                \hline
                $\chi^2$ &  & 2.93 &  &  & 3.65 &  &  & 2.54 & &  & 14.38 &  \\ 
                $p$ &  & 0.23 &  & & 0.16 &  &  & 0.28 &  &  & 0.001 &  \\ 
                \hline
        \end{tabular}
        \tablefoot{ F-I, F-II, F-I,II and F-III identify the different islands of solutions (see text). $q_{50}$ represents the median of the considered stellar quantities in the different solution islands, while $q_{16}$ and $q_{84}$ corresponds to the 16th and 84th quantiles. For some solutions, the distributions are skewed and $q_{16}$ or $q_{84}$ coincide with the median $q_{50}$. One asterisk marks the quantities, which differ by more than $1 \sigma$ from the corresponding observational constraints; two asterisks indicate a difference larger than $2 \sigma$. For each solution, the goodness-of-fit $\chi^2$ and the test $p$ value are reported.  }
\end{table*}

\begin{table*}[ht]
        \centering
        \caption{Multiple solutions for the  TZ~Fornacis binary system from the MESA grid of stellar models. Symbols have the same meaning as in Table~\ref{tab:sol-FRANEC}.}\label{tab:sol-MESA}
        \begin{tabular}{l|ccc|ccc|ccc}
                \hline\hline
                & \multicolumn{3}{c|}{M-I} & \multicolumn{3}{c|}{M-II} & \multicolumn{3}{c}{M-III}\\
                &  $q_{16}$ & $q_{50}$ & $q_{84}$ &  $q_{16}$ & $q_{50}$ & $q_{84}$ &  $q_{16}$ & $q_{50}$ & $q_{84}$    \\ 
                \hline
                $Y$ & 0.264 & 0.270 & 0.277 & 0.263 & 0.264 & 0.264 & 0.263 & 0.264 & 0.265 \\ 
                $Z$ & 0.013 & 0.014 & 0.016 & 0.014 & 0.015 & 0.016 & 0.014 & 0.016 & 0.016 \\ 
                $f_{\rm ov} $ & 0.012 & 0.013 & 0.014 & 0.023 & 0.023 & 0.024 & 0.027 & 0.027 & 0.028 \\ 
                age (Gyr) & 1.03 & 1.10 & 1.17 & 1.20 & 1.23 & 1.26 & 1.20 & 1.23 & 1.26 \\ 
                $M_{\rm cc}$ ($M_{\sun}$) & - & - & - & 0.143 & 0.144 & 0.146 & 0.164 & 0.166 & 0.167 \\ 
                \hline
                \multicolumn{10}{c}{Fit parameters}\\
                \hline
$T_{\rm eff,1}$ (K) &  & 4938 & & & 4915 & & & 4904 &  \\ 
$T_{\rm eff,2}$ (K) &  & 6824 & & & 6612 & & & 6708 &  \\ 
$R_1$ ($R_{\sun}$) &  & 8.24 & & & 8.52$^*$ & & & 8.61$^{**}$ &  \\ 
$R_2$ ($R_{\sun}$) &  & 4.02 & & & 3.79$^*$ & & & 3.69$^{**}$ &  \\ 
  \hline                
                $\chi^2$ &  & 2.44 &  &  & 7.93 &  &  & 18.11 &  \\ 
                $p$ &  & 0.30 &  &  & 0.02 &  &  &  < 0.001 &  \\ 
                \hline
        \end{tabular}
\end{table*}

The most probable class of solutions with the highest peak in the probability density at $\beta \approx$ 0.13 -- 0.15 for FRANEC (solutions F-I and F-II in Fig.~\ref{fig:franec_mesa-ov-age}) and $f_{\rm ov} \approx 0.013$ for MESA (solution M-I) suggests a primary star 
 in the central helium-burning phase and a secondary in the SGB phase. As expected, the inferred age and overshooting parameter values are highly correlated 
 as the larger the overshooting, the larger the convective core, and hence the longer the central hydrogen-burning phase. The 
 elongated shape of the first island of the probability density and its steep inclination suggest that a small variation of the convective core 
 overshooting has a significant impact on the inferred age. The comparison of the top-right end of this island of solutions 
 between the two panels of Fig.~\ref{fig:franec_mesa-ov-age} shows a minor difference, as the FRANEC grid provides a detached solution around $\beta \approx 0.17$ and 
 age 1.16 Gyr (F-II), whereas solutions based on a MESA grid do not show any gap. Tab.~\ref{tab:sol-FRANEC} and \ref{tab:sol-MESA} list the results 
 provided by FRANEC and MESA grid respectively, for the estimates of the binary system initial helium abundance, metallicity, age, core overshooting parameter and mass of 
 the convective core of the secondary star when present. 
 According to the multiple-solutions shown in Fig.~\ref{fig:franec_mesa-ov-age}, the tables separately report the results corresponding to the individual islands.
 The tables show the median values (i.e. $ q_{50} $) of the above mentioned quantities for the models selected by the maximum likelihood procedure described in Sect.~\ref{sec:method} and separated according to solution islands. The 16th ($q_{16}$) and 84th ($q_{84}$) quantiles are reported as representative of the $\pm 1 \sigma$ errors. The fitting values of temperatures and radii for the two stars are also shown. Finally, the tables report the goodness-of-fit $\chi^2$ of the obtained solutions and their $p$ values from the $\chi^2_2$ distribution. As usual, models with a goodness-of-fit $p$ value lower than 0.05 are considered as providing a poor agreement with observational constraints.

In detail, FRANEC grid solutions F-I and F-II have been computed by splitting the results at age 1.15 Gyr for $\beta$ < 0.20, while the MESA grid solution M-I 
 has been obtained for $f_{\rm ov} < 0.02$. The two FRANEC solutions differ slightly in terms of metallicity; F-I provides $Z$ = 0.013 and F-II $Z$ = 0.015.  
 Tab.~\ref{tab:sol-FRANEC} also shows the solution F-I,II for which the split in the two regions is neglected as the mentioned gap, rather than being genuine, could be an 
 artefact induced by the Monte Carlo procedure and/or by the metallicity resolution of the grid of models. Although consistent within the errors, the FRANEC FI,II solution 
 suggests both a lower initial helium abundance ($Y$ = 0.262) and a lower metallicity ($Z$ = 0.013) than the MESA M-I ($Y$ = 0.270, $Z$ = 0.014). However, both grids 
 provide results consistent with a helium-to-metal enrichment ratio of  $\Delta Y/\Delta Z = 1$. 
 The age of the system inferred by the two grids turned out to be nearly identical: 1.11$_{-0.03}^{+0.05}$ Gyr for F-I,II and 1.10 $\pm$ 0.07 for M-I. 
 
 In spite of the different numerical treatment of convective core overshooting in FRANEC and MESA codes, TZ~Fornacis binary seems able to 
 severely constrain the overshooting efficiency in both cases. The FRANEC $\beta$ parameter is in the range 0.14-0.16 with a best value of 0.155, while the MESA 
 $f_{\rm ov} $ is in the range 0.012-0.014 with median value 0.013. Such a result is in agreement with the qualitative relation $f_{\rm ov} \approx \beta/10$ 
 between the parameters adopted in a diffusive (as in MESA) and in an instantaneous (as in FRANEC) treatment of overshooting \citep{Noels2010}.
  
The goodness of the proposed fits can be judged relying on the $\chi^2$ as in Eq.~(\ref{eq:chi2}) summing over the two stars. Since there are six observational constraints (the masses are fixed to the observational value) and four free parameters (age, overshooting parameter, initial $Z$ and $Y$) in the fit algorithm, the computed statistics has two degrees of freedom. In Tab.~\ref{tab:sol-FRANEC} and \ref{tab:sol-MESA},  for both M-I and F-I,II  $p \approx 0.3$ and therefore one cannot reject the hypothesis that the solutions provide a good fit of the data.

\begin{figure*}
        \centering
        \includegraphics[width=8.3cm,angle=-90]{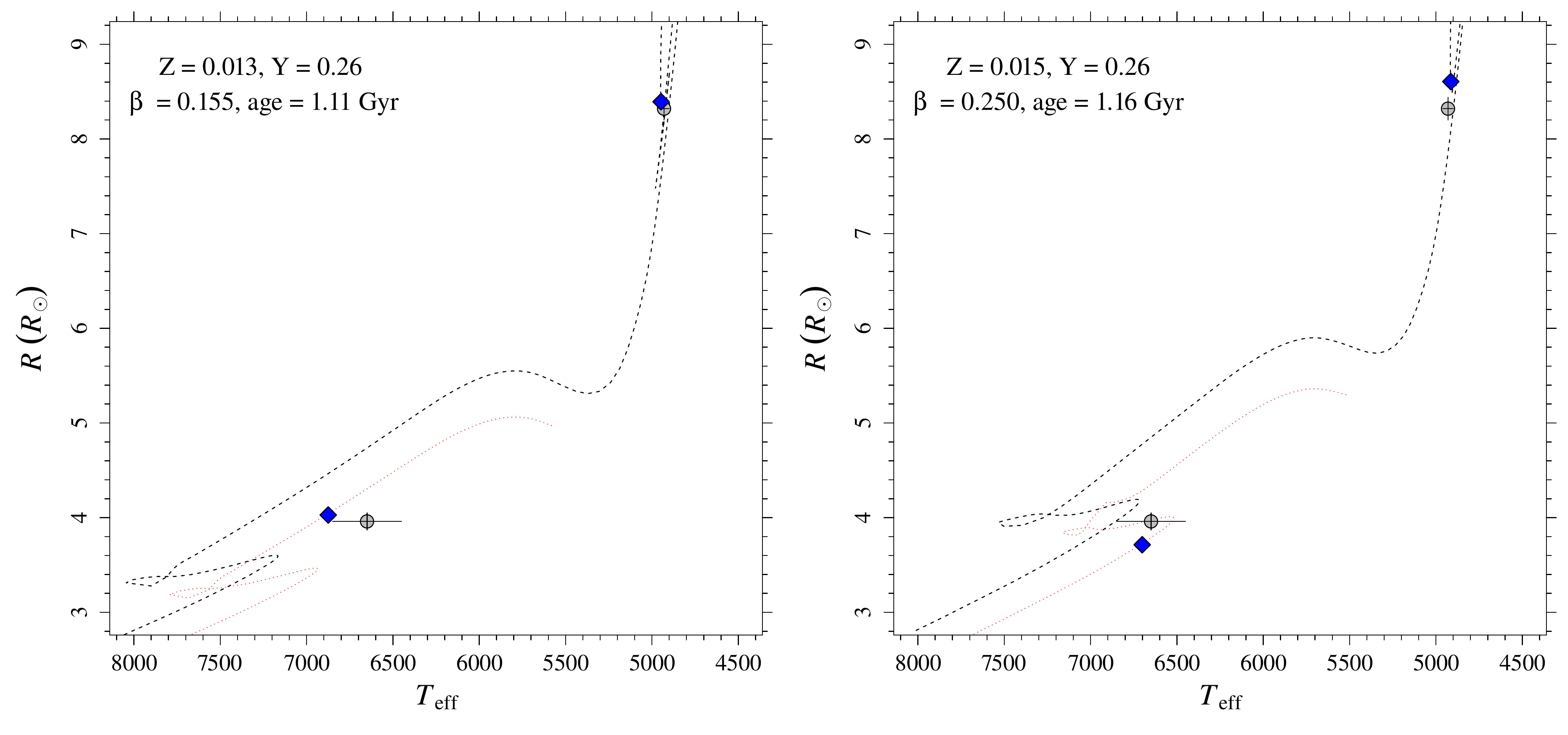}\\
        \includegraphics[width=8.3cm,angle=-90]{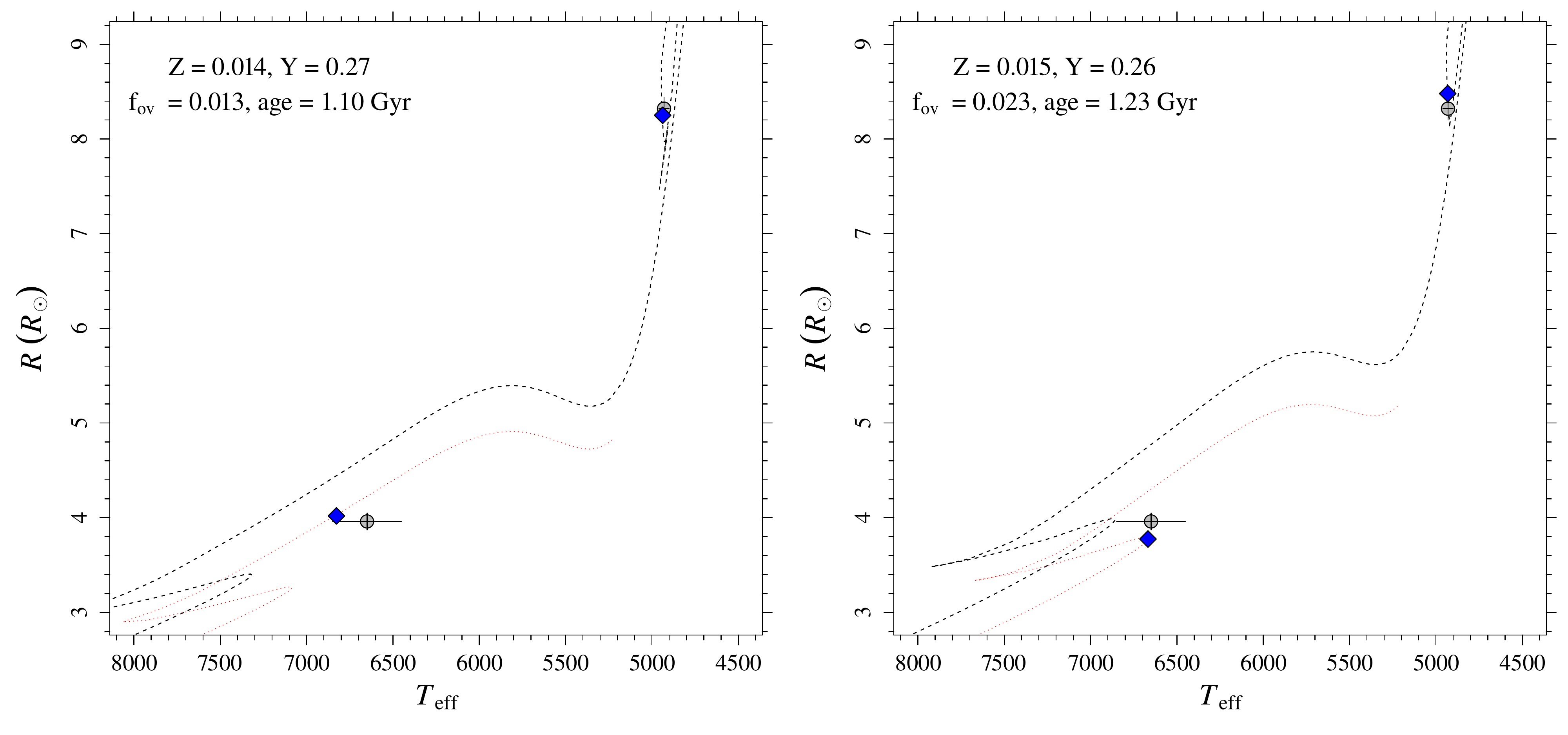}
        \caption{{\it Top row}: Comparison between the observational values of effective temperature and radius of the two stars (grey circle) and the evolutionary track for the best solutions found in the analysis of the FRANEC models. The blue diamonds mark the best fit theoretical positions for primary and secondary stars. The error bars correspond to 1 $\sigma$ errors. Initial metallicity and helium abundance, overshooting efficiency $\beta$, and age of the system for the four solutions are displayed in the panels. {\it Left panel} shows the solution F-I,II, while {\it right panel} shows solution F-III (see Tab.~\ref{tab:sol-FRANEC}). {\it Bottom row}: as in the {\it top row}, but  for tracks computed by MESA code.  {\it Left panel} shows the solution M-I, while {\it right panel} shows solution M-II (see Tab.~\ref{tab:sol-MESA}).}
        \label{fig:HR}
\end{figure*}

The FRANEC and MESA tracks in the plane effective temperature vs. radius for the solutions F-I,II and M-I are displayed in the left panels of Fig.~\ref{fig:HR}. The two evolutionary codes provide very consistent solutions by finding the secondary star in the SGB and the primary in the central helium-burning stage.

As anticipated at the beginning of this section, Fig.~\ref{fig:franec_mesa-ov-age} shows that the island of solutions described above is not unique and that 
there is another detached solution located at higher values of the overshooting parameter for both grids. However, the detailed morphology of such a second island 
of solutions differs in the two cases, since the FRANEC-based grid shows a single sharp peak while the MESA-based grid presents multiple and smoother peaks. 
For this reason, we analysed the MESA results by splitting the island into two sub-regions based on the value of the overshooting parameter $f_{\rm ov} = 0.025$. 
As in the case of the first island of solutions, the results based on the FRANEC grid are reported in Tab.~\ref{tab:sol-FRANEC} (case F-III) and those based on the MESA grid in Tab.~\ref{tab:sol-MESA} 
(cases M-II and M-III). Both codes suggest a slightly higher initial metallicity and an older age than in the 
first island of solutions. Concerning the metallicity, the agreement is perfect as F-III provides $Z$ = 0.015 and M-II and M-III $Z$ = 0.015 and $Z$ = 0.016, respectively. 
This is not the case for the age, as FRANEC F-III gives 1.16$_{-0.02}^{+0.03}$ Gyr while MESA gives $1.23 \pm 0.03$ Gyr for both M-II and M-III. The convective core overshooting efficiency 
is well constrained in both cases. The FRANEC $\beta$ parameter is in the range 0.24-0.255 with median value 0.25, while the MESA  $f_{\rm ov} $ is in the range 0.023-0.024 with median 0.023 
for M-II and 0.027-0.028 with median 0.027 for M-III. As in the first island of solutions, such results approximately follow the relation $f_{\rm ov} \approx \beta/10$. 
Concerning the evolutionary stage, this second class of solutions corresponds for both codes to a primary star in the central helium-burning phase, as in the first solution, 
but a secondary star in an earlier stage, that is, the overall contraction phase rather than in the SGB (see right panels in Fig.~\ref{fig:HR}). In this case, 
both grids of models fail to match the radii of the two stars and 
the solutions provide unsatisfactory fits, with high $\chi^2$. Even for the best of them (M-II), the $p$ value of the $\chi^2$ test is below the traditional acceptance level of 0.05, therefore leading us to reject the hypothesis that the recovered values provide a good fit to the data. 

As a final check, we performed an additional reconstruction relying on a MESA grid of stellar models computed by completely neglecting core overshooting in the central helium burning phase. The change is negligible and the only detectable variation is 
a decrease of 0.01 Gyr in the system age, well below the statistical error. The results are essentially independent of the details of the treatment of convective transport during 
central helium-burning because the primary star is still in an early stage in this phase. 

In summary, the maximum-likelihood procedure provides multiple solutions. According to a goodness-of-fit statistics, the best solution corresponds to a primary 
star in the central helium-burning phase and a secondary star in the SGB, regardless of the grid of adopted stellar models, namely F-I,II for FRANEC and 
M-I for MESA. 

It is interesting to note that the same difficulty in uniquely identifying a set of best fitting parameters has recently been described by \citet{Kirkby-Kent2016} for the AI~Phoenicis, a binary system composed of two stars of approximately 1.2 $M_{\sun}$. Relying on very high observational mass and radius accuracies  (0.4\% and 0.7\%, respectively), the evolutionary stage of the primary was firmly identified (the beginning of the RGB), while that of the secondary could not be unambiguously fixed. In fact, it was impossible to firmly place this star in the SGB or in the overall contraction phase, even if the space of the parameters adopted in the fit was much smaller than that used in the present analysis. In fact, the best fit models was established based on models at variable mixing-length values but fixed initial helium abundance and vice-versa. In all the cases the overshooting efficiency was keep fixed.
        
In the following section, we explore the effect of varying the observational errors on the inferred overshooting parameter and initial helium abundance. The impact of changing the prior constraints in the maximum likelihood procedure, such as keeping the $\Delta Y/\Delta Z$ ratio fixed and allowing different overshooting parameters, is discussed in Sect.~\ref{sec:ML-priors}.           

\section{Effect of observational uncertainties on parameter calibration}
\label{sec:errori-obs}

The previous analysis shows that even in a favourable case, such as TZ~Fornacis, the observational errors affecting the physical quantities used to constrain the maximum-likelihood procedure can hamper 
a definitive conclusion in regards to the age of the system and the core overshooting efficiency. It is therefore interesting to quantify the sensitivity of the 
recovered results on the observational errors. To do that, we performed further numerical experiments by repeating the maximum-likelihood procedure on the same data as in the previous cases, but this time this was done by artificially varying the observational precision.

\subsection{Effective temperatures and radii}
\label{sec:better-obs}

\begin{figure*}
        \centering
        \includegraphics[width=8.1cm,angle=-90]{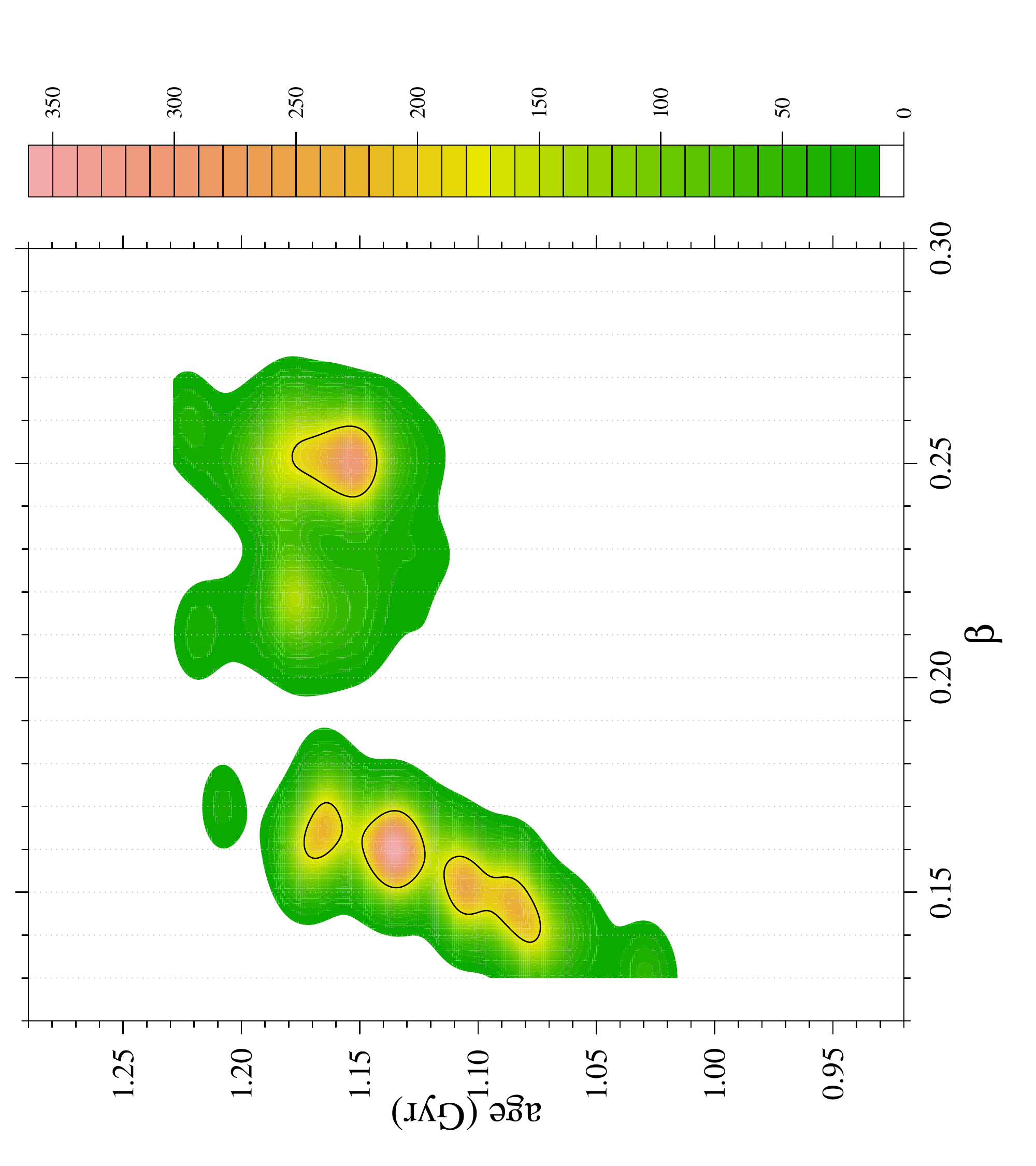}
        \includegraphics[width=8.1cm,angle=-90]{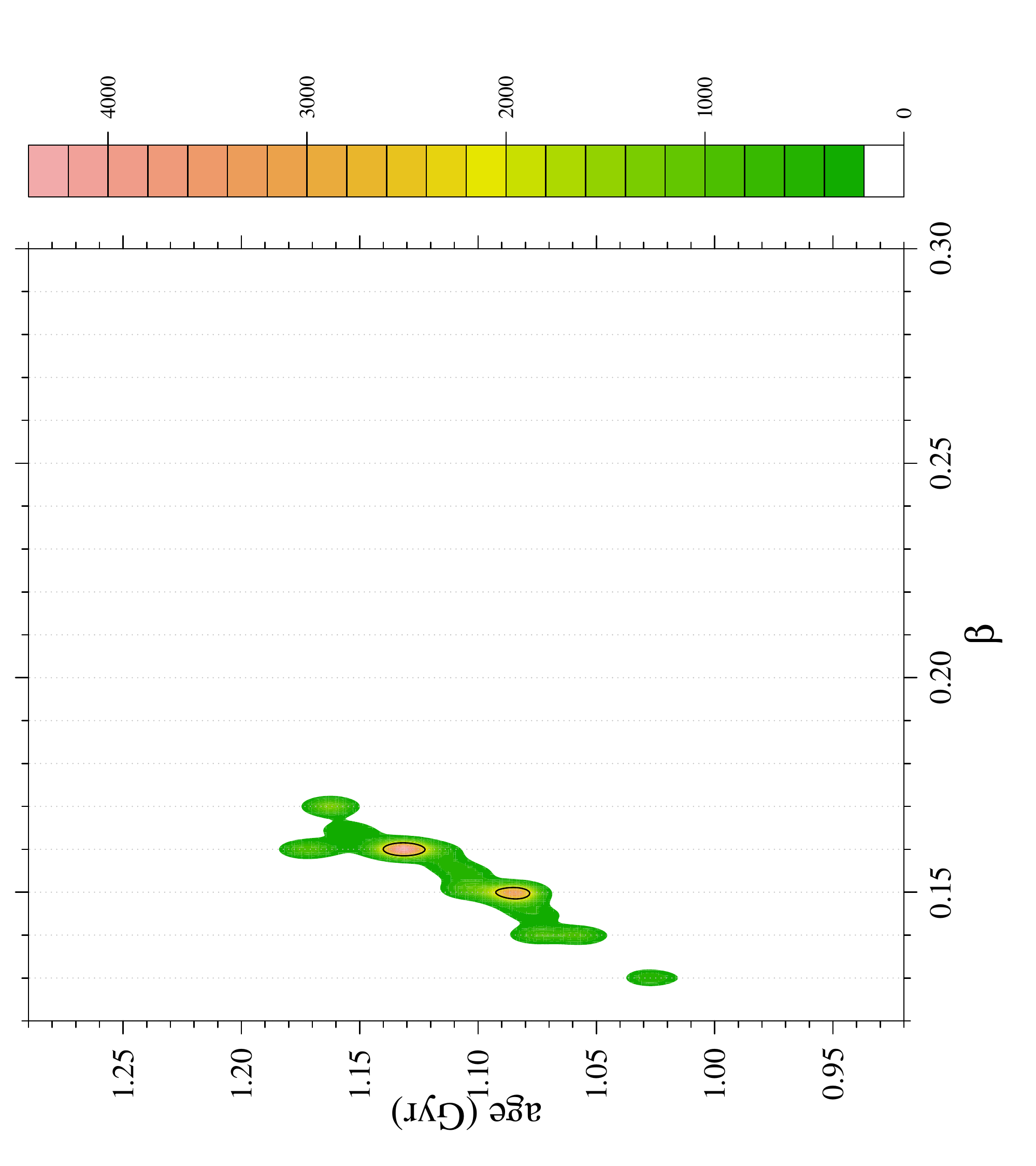}      
        \caption{{\it Left}: as in left panel of Fig.~\ref{fig:franec_mesa-ov-age}, but assuming an error of 100 K on the secondary effective temperature. {\it Right}: as in the left panel of Fig.~\ref{fig:franec_mesa-ov-age}, but halving the errors on the radii. The two panels were obtained by relying on the FRANEC grid of stellar models.}
        \label{fig:franec-halferr}
\end{figure*}

As a preliminary experiment, we halved
the uncertainty of the effective temperature of the secondary star (the effective temperature of the primary is already well constrained, see Tab.~\ref{tab:input}) and of the radii determinations. In particular, the relative error in the radii by \citet{Andersen1991} is of approximately 1.5\% for the primary star and 2.3\% for the secondary, while recent determinations for other binary systems often provide relative accuracy better than 1\%. 

Figure~\ref{fig:franec-halferr} shows the density of probability in the overshooting efficiency vs. age plane obtained from the FRANEC grid under the assumptions stated above. A comparison of the left panel of the figure with Fig.~\ref{fig:franec_mesa-ov-age} shows that the refinement in the secondary star temperature uncertainty plays a minor role. The low overshooting solution, identified by neglecting the fine structure for $\beta < 0.19$ as in Sect.~\ref{sec:results}, corresponds to $Z = 0.014 \pm 0.001$, $Y = 0.262_{-0.001}^{+0.003}$, $\beta = 0.160_{-0.015}^{+0.002}$ and age = $1.13_{-0.05}^{+0.03}$ Gyr. The $\chi^2$ of the solution is 3.60, with a $p$-value of 0.16.    
On the contrary, the increased precision in the radii produces a drastic change (right panel in Fig.~\ref{fig:franec-halferr}), since the solution at $\beta \approx 0.25$  vanishes, and the probability density shows two peaks around $\beta$ = 0.15 at $Z$ = 0.013 and 0.16 at $Z$ = 0.014. Both the two single solutions and the overall one, computed by neglecting the fine structure, provide satisfactory fits of the data, the lowest $p$-value being 0.11 for the $\beta = 0.15$ case.  

In summary, even whilst artificially increasing the data precision, the maximum likelihood algorithm based on a grid of models computed by allowing the initial helium content and overshooting efficiency to vary in reasonable uncertainty ranges is able to find a satisfying solution.  
However, owing to the continuous refinement of observational instruments and techniques, it can be expected that, in the near future, the same fitting procedure could result in no acceptable fit of the stellar data for a given system. Such a  result would allow us to test the general assumptions and limitations of the 1D stellar models.  

Another point to discuss is the very high accuracy in the $T_{\rm eff}$ of the primary star in the \citet{Gallenne2015} data. Given the difficulty to precisely evaluate the 
 systematic uncertainty, we repeated the analysis adopting a much more conservative error of 100 K for the primary star effective temperature. 
The results of the analysis are very similar to those of the reference scenario described in Sect.~\ref{sec:results}, with a slightly larger uncertainty. For the best fit solution: $\beta = 0.15^{+0.01}_{-0.02}$, $Y = 0.262 \pm 0.002$, $Z = 0.013 \pm 0.002$ and age = $1.09^{+0.07}_{-0.06}$ Gyr. The goodness of fit test for this solution is satisfactory ($\chi^2 = 2.59$, $p$ value = 0.27). A similar behaviour occurs for the second and unsatisfactory solution: $\beta = 0.250^{+0.005}_{-0.010}$, $Y = 0.263^{+0.001}_{-0.000}$, $Z = 0.015 \pm 0.001$ and age = $1.17 \pm 0.02$ Gyr ($\chi^2 = 14.06$, $p$ value = 0.001). The agreement of the results is not particularly surprising because, as can be seen in Table~\ref{tab:sol-FRANEC},  the constraint on the effective temperature of the primary star is always satisfied without problems; therefore, relaxing this constraint does not force  a significant change in the best fit solutions.

\subsection{Masses}
\label{sec:errorM}

It is well recognised in the literature that only very precise determination of stellar masses and radii can severely test stellar models \citep[see e.g.][]{Torres2010}. In particular, when attempting a calibration of a free parameter, as in the case of overshooting or mixing-length, 
a very high accuracy on fundamental parameters is mandatory, since uncertainties in the observational constraints will propagate in the final estimate. As discussed in the previous section, the unprecedented precision of mass 
determination of TZ~Fornacis members (at the level of 0.001 $M_{\sun}$) provided by \citet{Gallenne2015} allowed for calibration of several unknown parameters without further bothering of mass uncertainty. 

However, \citet{overshooting} recently showed that for binary systems where both stars are in the MS phase, uncertainties at the level of 1\% in mass determinations and of 0.5\% in radii,  values commonly found in current data,  
are high enough to hamper a statistically meaningful calibration of the core overshooting parameter. In spite of such evidence, it is unfortunately common to find overshooting calibrations in the literature  that 
fail to properly address the effect of the error in stellar mass estimates, even when uncertainties of the order of 1\% or larger are present. Such an assumption has no theoretical justification 
because the uncertainty on the calibrated parameter should be evaluated by propagating the errors in all the observational constraints. 
Moreover, the stellar mass is the most important parameter determining the evolution of the star; failing to address the uncertainty in its determination results in meaningless error on the final estimate.

The aim of this section is to demonstrate the importance of very accurate stellar mass measurements for calibration purposes. To do this, we again performed our maximum-likelihood procedure against TZ~Fornacis 
as above, but this time assuming larger values of the mass uncertainty. The comparison between these new results with those previously discussed will asses the impact of mass uncertainty on the calibration 
of stellar parameters. We performed such a numerical experiment twice and only for FRANEC models. In the first case, we adopted the most precise determination available in the literature
before the analysis by \citet{Gallenne2015} as error on primary and secondary masses,  that is, the values 
quoted in \citet{Andersen1991}: $\sigma(M_1) = 0.06$ $M_{\sun}$ (3\% relative error) and $\sigma(M_2) = 0.03$ $M_{\sun}$ (1.5\% relative error).  In the second case, we adopted a 1\% relative error on the two masses, a precision that is nowadays commonly achieved.  

In order to perform the recovery procedure in presence of mass uncertainty of this order of magnitude, the grid of pre-computed stellar models must be extended 
to properly account for the larger range of variation, as explained in Sect.~\ref{sec:grids}.  

\begin{figure*}
        \centering
        \includegraphics[width=8.1cm,angle=-90]{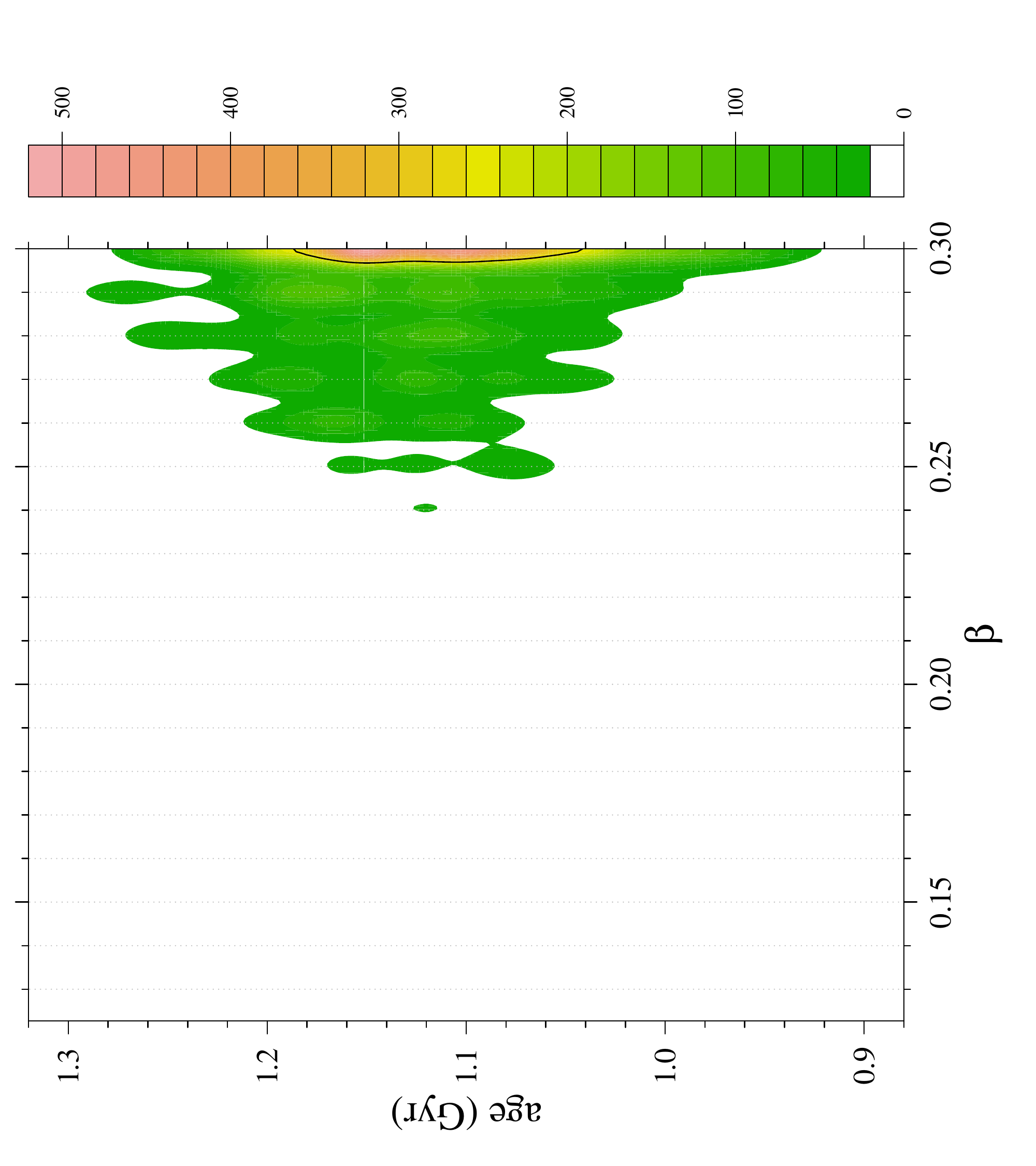}
        \includegraphics[width=8.1cm,angle=-90]{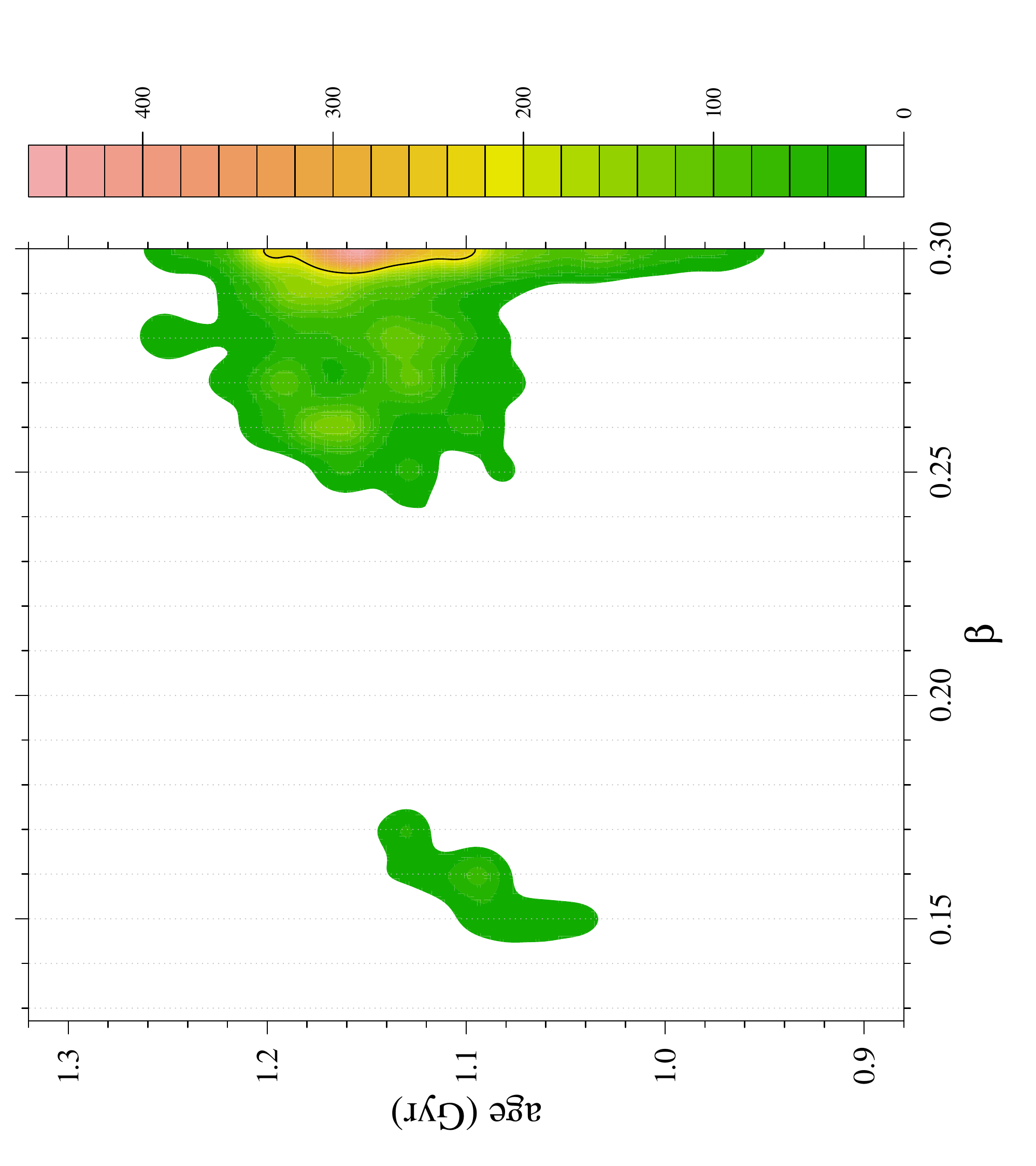}
        \caption{{\it Left}: as in Fig.~\ref{fig:franec_mesa-ov-age}, but for estimates obtained while assuming errors of 1.5\% and 3\% on the primary and secondary star masses. A multi-mass grid of stellar models was adopted for the recovery (see text). {\it Right}: as in the {\it left} panel, but assuming errors of 1\% on both stellar masses. }
        \label{fig:franec-ov-age-multimass}
\end{figure*}

Figure~\ref{fig:franec-ov-age-multimass} shows the 2D density of probability of the estimated core overshooting parameter $\beta$ and age for the two different assumptions about the mass uncertainty. 
The best-fit stellar parameters and their errors for both scenarios are collected in Tab.~\ref{tab:sol-FRANEC-MM-He}. The MM-A solution refers to the case of mass uncertainty from \citet{Andersen1991}, while MM-B refers to the case of 1\% uncertainty on stellar masses. 

\begin{table*}[ht]
        \centering
        \caption{Multi-mass and fixed initial helium solutions from FRANEC grids of stellar models. Symbols have the same meaning as in Table~\ref{tab:sol-FRANEC}.}\label{tab:sol-FRANEC-MM-He}
        \begin{tabular}{l|ccc|ccc|ccc}
                \hline\hline
                
                & \multicolumn{3}{c|}{MM-A} & \multicolumn{3}{c|}{MM-B} & \multicolumn{3}{c}{$\Delta Y/\Delta Z = 2$}\\
                &  $q_{16}$ & $q_{50}$ & $q_{84}$ &  $q_{16}$ & $q_{50}$ & $q_{84}$ &  $q_{16}$ & $q_{50}$ & $q_{84}$    \\ 
                \hline
                $Y$ & 0.262 & 0.263 & 0.271 & 0.262 & 0.263 & 0.266 & 0.272 & 0.275 & 0.277\\ 
                $Z$ & 0.013 & 0.014 & 0.015 & 0.013 & 0.014 & 0.016 & 0.012 & 0.013 & 0.014\\ 
                $\beta$ & 0.270 & 0.300 & 0.300 & 0.270 & 0.297 & 0.300 & 0.120 & 0.130 & 0.130\\ 
                age (Gyr)& 1.04 & 1.12 & 1.19 & 1.09 & 1.15 & 1.19 & 0.97 & 1.01 & 1.02\\ 
                $M_{\rm cc}$ ($M_{\sun}$) & 0.169 & 0.174 & 0.180 & 0.169 & 0.172 & 0.175 & - &-&-\\ 
                \hline
                $\chi^2$ &  & 3.34 &  &  & 2.74 & &  & 8.08 &\\ 
                $p$ &  & 0.50 &  &  & 0.60 & &  & 0.04 & \\ 
                \hline
        \end{tabular}
\end{table*}

The striking difference between the two panels of Fig.~\ref{fig:franec-ov-age-multimass} and the left panel of  Fig.~\ref{fig:franec_mesa-ov-age} is due solely to the difference in the mass uncertainty  used in the 
maximum-likelihood procedure. A simple increase of the mass uncertainty completely alters the morphology of the best-fit solution. Concerning the calibration of the overshooting, the less precise is the mass and the more  skewed 
toward the largest value present in the grid of models is the estimated $\beta$ parameter, with best fit $\beta = 0.3$ for both scenarios. In the case of the largest mass error (3\%, left panel), the island of solutions at low overshooting values, which was the most probable one in 
Fig.~\ref{fig:franec_mesa-ov-age}, is completely missing, while it begins to manifest itself, even if at low density of probability, in the case of mass error at 1\% level. 
The $p$ values of the goodness of fit tests for these solutions are higher than those reported in Tab.~\ref{tab:sol-FRANEC}, and therefore the solutions are legitimate. Such a behaviour is expected because the fitting algorithm can explore a larger range of mass combinations.

\begin{figure*}
        \centering
        \includegraphics[width=8.1cm,angle=-90]{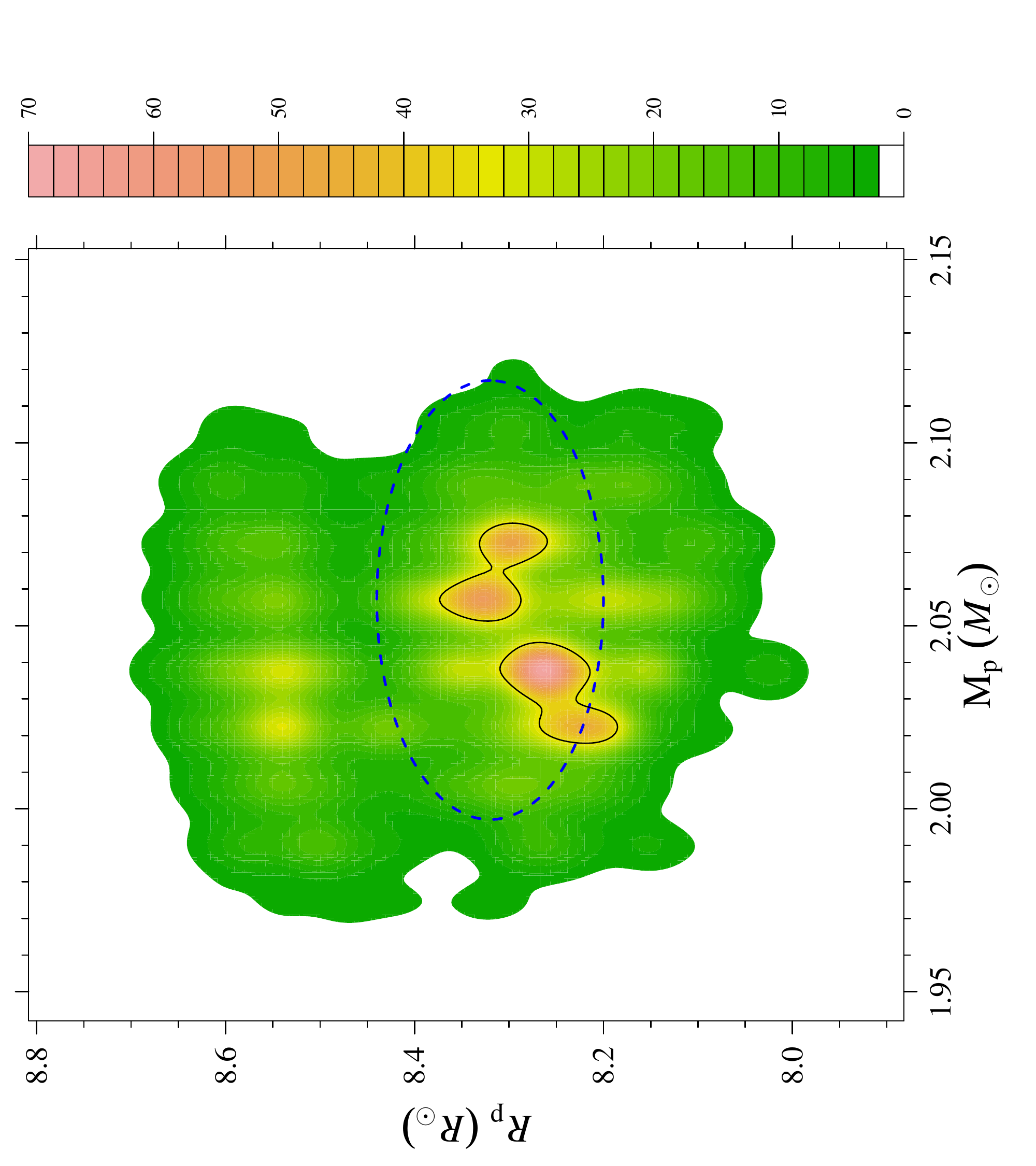}
        \includegraphics[width=8.1cm,angle=-90]{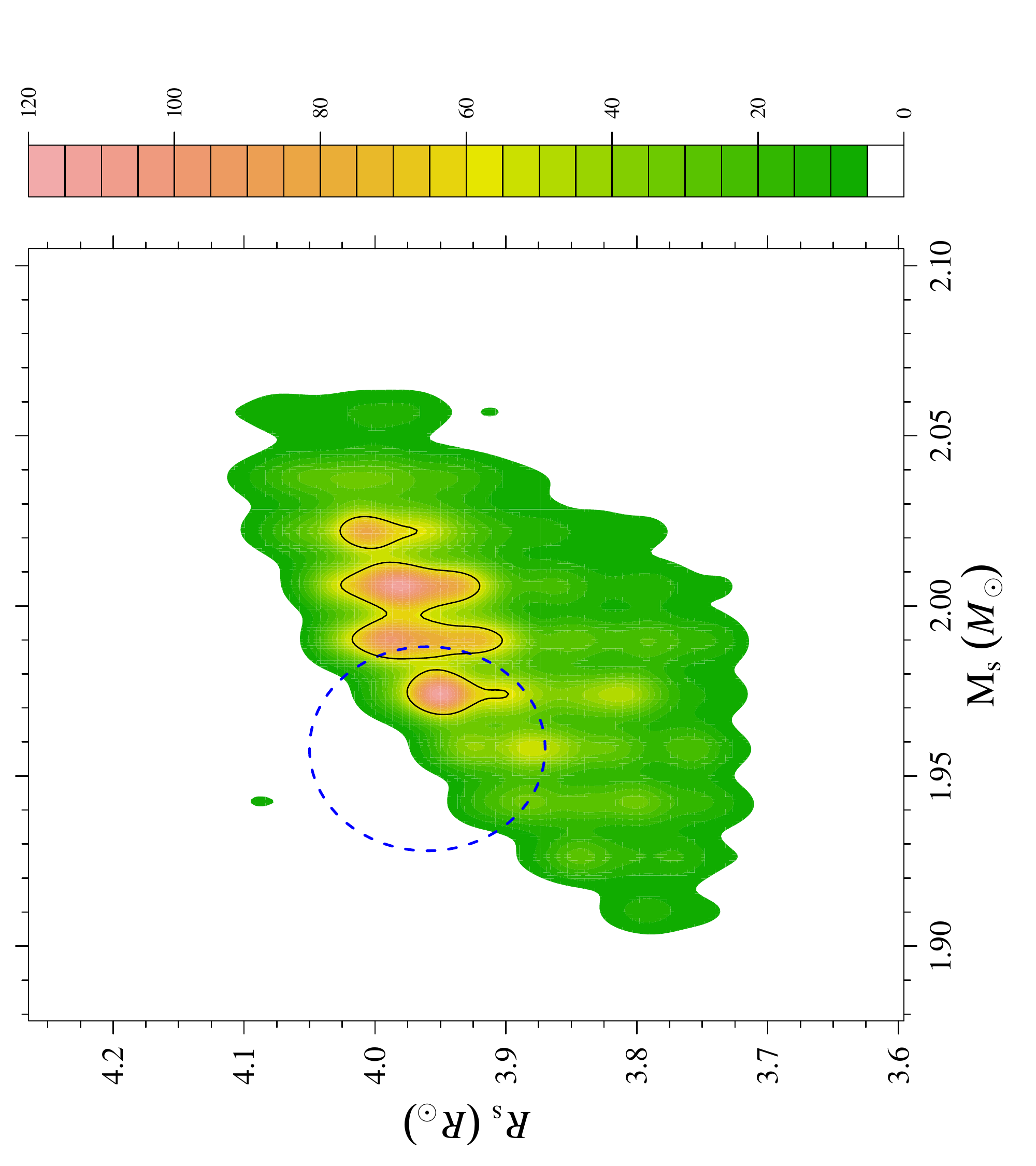}        
        \caption{({\it Left}): 2D density of probability for the best-fit mass and radius for the primary star. The blue-dashed ellipse marks the 1 $\sigma$ uncertainty on the observational values. The results were obtained relying on a multi-mass grid, with errors on the stellar masses from \citet{Andersen1991} (see text). {\it Right}: same as in the {\it left} panel, but for the secondary star.}
        \label{fig:franec-MR-multimass}
\end{figure*}

It is thus apparent that an error of a few percent on the mass determination can not only broaden the uncertainty on the estimated parameters, but, as in this case, can also drastically affect the best fit solutions. 

The explanation of this behaviour merits a brief analysis. It is clear that the 
larger the uncertainty affecting the observational constraints, the
wider the parameter space that can be explored by the fit algorithm. Indeed, a fit of a system will never be perfect due to both the discrepancies between synthetic and real single stars and the modelling of the evolution of a binary system. Even in an ideal world in which the two stars evolve independently and theoretical models perfectly reproduce the data, a simple random error in the observations will produce a mismatch between fitted and real objects (see \citealp{overshooting} for an example of the possible magnitude of such an effect).
Hence, it is reasonable to expect that the result obtained under tighter observational constraints provides a worse fit than that obtained under looser ones. This is particularly true where the mass of the stars is concerned because of its major effect on stellar evolution.      
The key point we would highlight here is that the possibility of testing the assumptions of stellar evolution in binary systems requires an accuracy in mass determination much higher than 1\%, otherwise the propagation of larger mass uncertainty allows the algorithm to explore an overly wide range of mass combinations. This additional variability can mask a possible slight difference between theory and observations, since it is easier for the algorithm to find a track combination compatible with the observations.
Although the correlation imposed in the covariance matrix between the masses  (see Sect.~\ref{sec:method}) protects against  unphysical inversion of the mass ratio, the algorithm is free to adjust the mass inside the nominal observational errors. To illustrate this effect, Fig.~\ref{fig:franec-MR-multimass} shows 
the density of probability of reconstructed masses and radii for both primary (left panel) and secondary (right panel) stars, computed from the largest mass error set. From the right panel of the figure, it is clear that the mass of the secondary star is preferentially overestimated with respect to the observational value.  
This overestimation occurred to reduce the tension between the observational constraints and the grid of theoretical models. In detail, the shift in the fitted mass of the secondary is of approximately 0.04 $M_{\sun}$, a value large enough to heavily influence the theoretical stellar track. It is therefore unsurprising that the calibrated age and overshooting of this system are completely different from those described in Sect.~\ref{sec:results}. The islands of solutions present in Fig.~\ref{fig:franec_mesa-ov-age} no longer appear in the left panel of Fig.~\ref{fig:franec-MR-multimass} because their likelihood is dominated by that of models at different masses.

\section{Effect of different assumptions in the maximum-likelihood fit}
\label{sec:ML-priors}

\subsection{Keeping the helium-to-metal enrichment ratio fixed}
\label{sec:dydz}

Stars such as TZ~Fornacis members are too cold for spectroscopical constraint of their helium content, resulting in an additional unknown parameter to be calibrated,
as the initial helium abundance significantly affects the evolution of stars of given mass and metallicity. Stellar modellers usually assume
a linear relationship between the initial metallicity and helium abundance, that is, $Y = Y_p+\frac{\Delta Y}{\Delta Z} Z$. However, the $\Delta Y/\Delta Z$ ratio is still
poorly constrained \citep[e.g.][]{pagel98,jimenez03,gennaro10} and consequently the initial helium abundance to be
adopted in stellar models at a given metallicity is subject to some variability. For this reason, we computed a grid of models
for five values of $\Delta Y/\Delta Z$ in the range 1-3 (see Sect.~\ref{sec:grids}) and 
we treated the helium-to-metal enrichment ratio as a parameter to be estimated by the maximum-likelihood procedure, as is the case for the overshooting and the age.

However, in the literature, it is much more common to find grids of models computed at fixed $\Delta Y/\Delta Z$ ratios, as is the
case of publicly available databases, where a value of approximately two is usually adopted. Therefore, it is worthwhile
analysing the effect of adopting
a grid of stellar models computed by keeping the helium-to-metal enrichment
ratio fixed on overshooting calibration and age estimate.

\begin{figure}
        \centering
        \includegraphics[width=8.1cm,angle=-90]{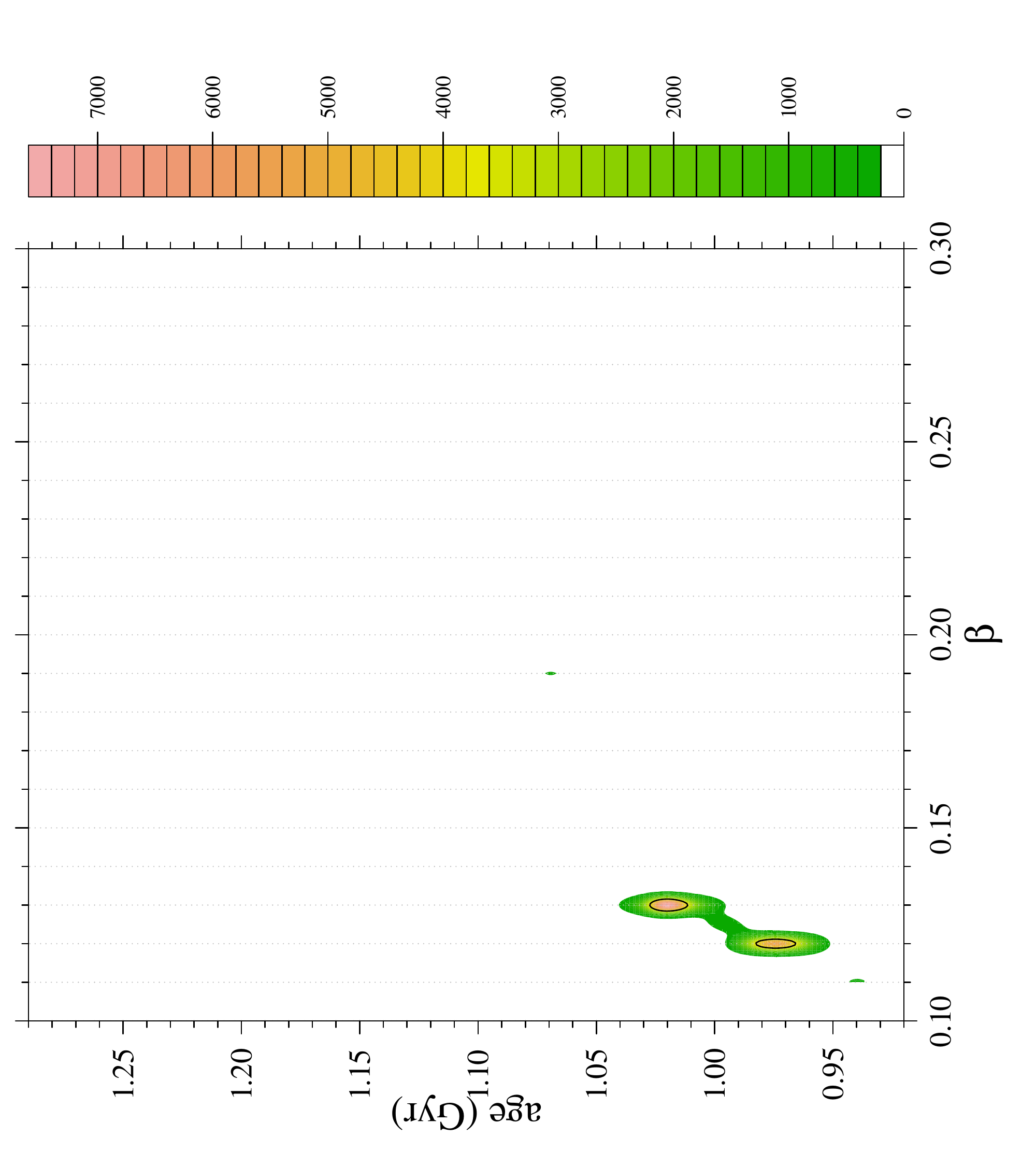}       
        \caption{Same as in Fig.~\ref{fig:franec_mesa-ov-age} for FRANEC grid, but for estimates obtained by keeping fixed the helium-to-metal enrichment ratio $\Delta Y/\Delta Z = 2$.}
        \label{fig:franec-dydz2}
\end{figure}

Figure~\ref{fig:franec-dydz2} shows the density of probability in the overshooting efficiency vs. age plane obtained from the FRANEC grid but only by models with $\Delta Y/\Delta Z = 2$.  The impressive
difference between this figure and the left panel of Fig.~\ref{fig:franec_mesa-ov-age}  is due solely to the different assumption on the
helium-to metal enrichment ratio (fixed vs. variable), as all the other characteristics of the grid of stellar models are exactly the same.

Figure~\ref{fig:franec-dydz2} shows two distinct peaks at $\beta = 0.12$ and 0.13, but, as stated above, this sub-structure could possibly  be due to the grid step in the overshooting parameters. The stellar parameters inferred from the fit, neglecting the aforementioned sub-structure, are presented in Tab.~\ref{tab:sol-FRANEC-MM-He}, labelled as $\Delta Y/\Delta Z=2$.
As can be expected, Fig.~\ref{fig:franec-dydz2} shows that forcing the estimates to adopt a higher value of initial helium content than those preferred by the maximum likelihood solution in Sect.~\ref{sec:results} (0.275 vs. 0.262) shifts the age and the overshooting parameter estimates towards lower values, the system age being in the range [0.97; 1.02] Gyr. 
Moreover, the density of probability is highly constrained in a small region of the overshooting efficiency vs. age plane because only a very small subset of models computed adopting $\Delta Y/\Delta Z = 2$ is compatible with the observational data.  
The goodness-of-fit $\chi^2$ test -- with three degree of freedom, since the initial helium content is no longer estimated from the data -- for the solution that neglects the peak sub-structure gives a $p$-value of 0.04, and therefore a poor fit of the data.  
This is an expected behaviour, because the fitting algorithm is forced to explore a reduced space of parameters, away from the best-fitting region found in Sect.~\ref{sec:results}, for determining the solution of the system.

In conclusion, neglecting the current uncertainty in the initial helium abundance and adopting stellar models computed at a fixed $\Delta Y/\Delta Z$
severely affects both the age estimate and the overshooting calibration, confirming what has already been shown by \citet{overshooting}  for eclipsing binaries whose members are in MS. 

\subsection{Allowing different overshooting efficiency for the two stars}\label{sec:diff-ov}

As anticipated in Sect.~\ref{sec:intro}, the choice to fit the system adopting a common value of overshooting efficiency is based on the theoretical reference framework of independent evolution of the two stars.  Under this assumption, with the two stars having almost the same mass,  the possibility of different overshooting parameters in the two binary members can be safely neglected. However, it is interesting to discuss the methods for testing the need for different overshooting efficiency in the two stars and some shortcomings of this approach.          
        
Although the solutions with low overshooting efficiency from FRANEC and MESA grids proved to be satisfactory in the goodness of fit test, a much better solution with independent overshooting efficiency for the two stars could, in principle, exist. To test this hypothesis, we repeated the estimation process relaxing the constraint of a common $\beta$ in the FRANEC grid, therefore losing one degree of freedom in the $\chi^2$ test. 
        
        \begin{table}[ht]
                \centering
                \caption{Solutions obtained whilst relaxing the common overshooting efficiencies  hypothesis from FRANEC grids of stellar models. Symbols have the same meaning as in Table~\ref{tab:sol-FRANEC}.}\label{tab:sol-OV}
                \begin{tabular}{l|ccc|ccc}
                        \hline\hline
                        
                        & \multicolumn{3}{c|}{OV-I} & \multicolumn{3}{c}{OV-II} \\
                        &  $q_{16}$ & $q_{50}$ & $q_{84}$ &  $q_{16}$ & $q_{50}$ & $q_{84}$ \\ 
                        \hline  
                        $Y$ & 0.261 & 0.263 & 0.269 & 0.262 & 0.263 & 0.264 \\ 
                        $Z$ & 0.012 & 0.013 & 0.015 & 0.013 & 0.015 & 0.016 \\ 
                        $\beta_1$ & 0.140 & 0.151 & 0.160 & 0.160 & 0.180 & 0.200 \\ 
                        $\beta_2$ & 0.030 & 0.058 & 0.080 & 0.260 & 0.270 & 0.270 \\ 
                        age (Gyr) & 0.96 & 1.02 & 1.06 & 1.15 & 1.18 & 1.20 \\ 
                        $M_{\rm cc}$ ($M_{\sun}$) & - & - & - & 0.169 & 0.171 & 0.173 \\ 
                        \hline
                        $T_{\rm eff,1}$ (K) &  & 4937 & & & 4922 &  \\ 
                        $T_{\rm eff,2}$ (K) &  & 6779 & & & 6667 &  \\ 
                        $R_1$ ($R_{\sun}$) &  & 8.27 & & & 8.57$^{**}$ &  \\ 
                        $R_2$ ($R_{\sun}$) &  & 4.04 & & & 3.78$^{**}$ &  \\ 
                        \hline
                        $\chi^2$ &  & 1.42 &  &  & 8.96 &  \\ 
                        $p$ &  & 0.23 &  &  & 0.003 &  \\ 
                        \hline
                \end{tabular}
        \end{table}
        
The results of the analysis are presented in Table~\ref{tab:sol-OV}. For the solution at low overshooting efficiency, the secondary star is fitted with $\beta_2 = 0.058^{+0.022}_{-0.028}$, a much lower value than the primary $\beta_1 = 0.151^{+0.009}_{-0.011}$. At first glance, it seems, therefore, that a different overshooting efficiency would be suggested for the two stars. 
However, such a conclusion is not statistically significant since it does not rely on a rigorous test. 
        
The proper way to statistically evaluate the importance of additional parameters inserted in a model is based on the comparison of nested models, a ubiquitus statistical method of inference \citep[see e.g.][]{venables2002modern,simar}. This approach relies on the comparison of appropriate goodness of fit statistics for two nested models (in this case the fit $\chi^2$). 
The two models we are interested in comparing are nested because they rely on the same data and grids, but they differ since the model with independent overshooting contains one additional parameter to be estimated. 
To follow the statistical nomenclature, let us call the model that allows for different overshooting efficiency between the two stars the {\it full} model, and the alternative one, the {\it restricted} model. Let $\nu_{\rm f}$ and  $\nu_{\rm r}$ be the degrees of freedom of the full and  the restricted models, respectively. Then the statistic $G = \chi^2_{\rm r} - \chi^2_{\rm f}$ has a $\chi^2(\nu_{\rm r}-\nu_{\rm f})$ distribution, allowing us to test the statistical significance of additional parameters in the model. In our case, for the comparison of models OV-II and F-I,II, this gives $G =  2.54 - 1.42 = 1.10$ with one degree of freedom. The $p$ value of the test is 0.29, therefore the two models do not differ significantly, and offer an equivalent fit of the data. 
Hence, we cannot conclude that the solution with different overshooting is better than the other with a single value;
relying on the principle of maximum parsimony, the model that assumes a common overshooting efficiency is therefore preferable.
        
In addition to all these statistical considerations,  
allowing the stars to have independent overshooting efficiencies would lead to an additional free parameter, which can easily mask the possible difficulty of stellar tracks to fit the system. In other words, the overshooting efficiency would reflect not only the additional mixing of the core, but also other effects from any given source, which would remain hidden. 
Therefore, although a difference in the overshooting parameter can be theoretically required and justified when the two stars have different masses, it is somewhat questionable when they haven't. 
Moreover, the recent theoretical works by \citet{testW}  and \citet{overshooting} show, in  detail, the variability in the difference of age and overshooting efficiency that can be recovered by stars in binary systems with both components in MS phase. These works  show that, even in the overoptimistic case when stars and models are in perfect agreement, that is, when the artificial stars are sampled from the same theoretical tracks adopted in the recovery, 
the variability in the recovered parameters is large. As an example, Fig.~4 of \citet{overshooting}  shows that for two stars in MS phase, and assuming an uncertainty of 1\% on the masses and 0.5\% on the radii, a difference of approximately 0.15 in the recovered overshooting parameter $\beta$ can be expected to occur owing solely to the statistical uncertainty in the observational constraints. Although these results cannot be directly adopted for the present study, they should nonetheless be considered before claiming a significant
difference in the individually recovered overshooting efficiencies of the two stars. More thorough theoretical investigations are needed before relying on calibrations that could be the results of random fluctuations in the observables.

\section{Comparison with the literature}\label{sec:literature}

Several age estimates  for TZ~Fornacis exist in the literature, such as $1.75 \pm 0.15$ Gyr \citep{Andersen1991}; 1.3 Gyr \citep{Claret1995}; [1.12; 1.33] Gyr \citep{VandenBerg2006} and [1.08; 1.41] Gyr \citep{Claret2011}. Apart from the first determination, which is based on outdated opacity tables, the others agree relatively well within each other and are consistent with our best estimate. 
The overshooting efficiency for TZ~Fornacis was estimated by \citet{Stancliffe2015} who, adopting MESA, found $f_{\rm ov} = 0.05$, a value much larger than our estimates. The difference can be partially ascribed to differences in the fitting algorithms, but mainly in the explored evolutionary phases; the grid employed in that paper does not include models in the central helium burning stage,  our preferred solution, and to the initial helium values adopted in the grids.

However, the most relevant comparison is the one with the results presented by \citet{Gallenne2015}, who adopt their newly determined mass and uncertainties. They provided a system age of $1.2 \pm 0.1 $ Gyr and placed the secondary stars in the contraction phase by using stellar evolutionary tracks from BASTI \citep{teramo04} and PARSEC \citep{Bressan2012}, which adopt  $\beta$ = 0.20 and 0.25, respectively. Therefore, the solution identified in that paper is most similar to the one highly disfavoured by our analysis.

It is also interesting to compare the reconstructed best-fit overshooting values with results coming from asteroseismic observations of less massive stars. Recently,
by relying on  the detections of
{\it Kepler} \citep{Borucki2010}, \citet{Deheuvels2016}  estimated
the extent of the convective core in eight MS stars in the mass range [1.12-1.45] $M_{\sun}$. The resulting overshoot parameters, assuming instantaneous overshooting as in FRANEC code, were between 0.05 and 0.15 (with typical uncertainty of $\pm 0.01$) when microscopic diffusion is included in the computations. A possible increasing trend of overshoot parameter with stellar mass was also demonstrated, but much more data are required to definitively confirm it \citep[see also the results from][from eclipsing binaries]{Claret2007,Claret2016}.
However, it should be noted that the definition of the overshooting extension beyond the classical Schwarzschild border adopted in  \citet{Deheuvels2016} is somewhat different from those used here, and that the calibrated value of the overshooting parameter depends on the assumed chemical and input physics  in the stellar computations. Therefore a much more meaningful comparison would concern the mass extension of the convective core rather than the overshooting parameter.

\section{Conclusions}\label{sec:conclusions}

Taking advantage of the very precise mass determination of the stars in the eclipsing binary system TZ~Fornacis recently provided by \citet{Gallenne2015}, we tried to constrain the convective core overshooting, the initial helium abundance and the age of the system. 
We assumed independent evolution of the two stars and thus, having the binary members nearly equal in mass, the same overshooting efficiency for both components.
We put particular effort into the statistical treatment of both the parameter estimates and of their errors. To do this, we used the SCEPtER pipeline \citep{scepter1,eta,binary}, a maximum likelihood procedure relying on several dense grids of stellar models.
To check the possible systematic effects on the results of the adopted stellar evolutionary code, we computed two grids of models by means of independent codes, namely FRANEC and MESA. 
In addition to the best fit, we also discussed the impact on the calibration results of varying the observational uncertainties, of keeping fixed the $\Delta Y/\Delta Z$ ratio, and of allowing different overshooting efficiencies for the two binary members.

Concerning the best fit, overall, the results from within these two grids agree relatively well with each other.
Despite the very high quality of the data, with errors in mass lower than 0.1\% and in radius of between 1.5 and 2.2\%, the quoted stellar parameters cannot be unambiguously constrained. In fact, regardless of the grid of stellar models used, we found 
at least two distinct classes of solutions. 

The primary star is in the central helium-burning phase in both classes of solutions, whereas the secondary star is in the SGB phase in the first class and in the overall-contraction phase at the central 
hydrogen exhaustion in the second class. 

Concerning the first class of solutions, the best fit based on FRANEC models provides an age of $1.11_{-0.03}^{+0.05}$ Gyr, a core overshooting parameter $\beta = 0.15 \pm 0.01$ and an initial helium abundance $Y= 0.262 \pm 0.001$. These values are in 
agreement with those based on MESA models, that is, age $1.10 \pm 0.07$ Gyr, diffusive overshooting $f_{\rm ov}= 0.013 \pm 0.001 $ and $Y= 0.270_{-0.006}^{+0.007} $. Both grids of models suggest a low helium-to-metal enrichment ratio $\Delta Y/\Delta Z = 1$.

The second class of solutions yields a more efficient convective core overshooting both for the FRANEC $\beta = 0.25_{-0.01}^{+0.005}$ and MESA $f_{\rm ov} = 0.025 \pm 0.003$ formalism. In this case, the estimated ages are $1.16_{-0.02}^{+0.03}$ Gyr and 
 $1.23 \pm 0.03$ Gyr, respectively. However, the goodness-of-fit of this second class of solutions is much worse than that of the first class. 
  
Besides the determination of the best fit values of age and overshooting parameter for the TZ~Fornacis system, in this work we specifically addressed some methodological and statistical problems affecting the calibration of free parameters from binary systems. We  investigated how different yet legitimate choices  in the computation of the stellar model grid and in the error treatment can produce completely different calibrations. 
  
In particular, we showed the importance for calibration purposes of relying on very accurate stellar mass determination. To this aim, we repeated the estimation process with identical input but assuming larger yet still relatively typical uncertainties in the two stellar masses. As a result, we obtained that even an uncertainty at a level of 1\% causes a severe bias in the inferred overshooting parameter, completely altering the morphology of the best fit solution.
   
We also studied the effect of neglecting the uncertainty in the helium-to-metal enrichment ratio, and thus of the initial helium content, on the stellar parameter estimates. To do this, we repeated the fitting procedure relying solely on models with fixed $\Delta Y/\Delta Z = 2$, a value frequently adopted in the literature.
The result shows a notable bias, larger than the statistical error, in the estimates of system age and overshooting efficiency toward lower values.

The possibility of a different overshooting parameter for the two stars was also investigated, even if having two
stars of nearly identical mass is theoretically unlikely. However, this choice did not improve the quality of the fit, and can lead to artefacts in the solutions. 
        
In summary, we tried to show that the calibration of parameters from binary systems is a delicate task, affected by many decisions in the fitting stage. One of the main results of the analysis is the quantification of the importance to properly account for various sources of uncertainty that affect both the observational data and the theoretical model computations. 
Clearly this means a heavy computational effort because it requires the building of very large and fine grids of stellar models. However, it is a mandatory effort since fixing some values in the theoretical computation without observational evidence (i.e. the initial helium content) or neglecting the actual error in the masses of the two stars can
severely affect the calibration from observations. On the other hand, this study provides a warning against relying on fit, which allows for too much variability, such as allowing for independent overshooting efficiency for stars of nearly equal mass, because the calibration can be significantly affected by mere random fluctuations. 

Although we explicitly address  many sources of uncertainty, we did not evaluate the possible systematic effects of fixing the input physics  in stellar model computations. The quantification of this uncertainty would require a huge amount of computation and could not be explored in the present work.

More theoretical investigations aimed at quantifying the variability expected
by chance of parameters calibrated from binary systems is needed before relying on the results obtained in this way. The continuously growing availability of powerful computation facilities for calculating theoretical stellar models under a wide range of input, and the development of sound statistical techniques for obtaining and comparing the best fit solutions, will offer a unique opportunity to firmly evaluate the actual  degree of reliability of these calibrations.

\begin{acknowledgements}
We thank our anonymous referee for many stimulating suggestions that
helped to clarify and improve the paper.
This work has been supported by PRIN-INAF 2012 ({\em The M4 Core Project with Hubble Space Telescope}, PI
L. Bedin ), PRIN-INAF 2014 (\emph{The kaleidoscope of stellar populations in globular clusters with Hubble Space Telescope}, PI S. Cassisi), PRA Universit\`{a} di Pisa 2016 
(\emph{Stelle di piccola massa: le pietre miliari dell'archeologia galattica}, PI: S. Degl'Innocenti) and by INFN (\emph{Iniziativa specifica TAsP}).
\end{acknowledgements}

\bibliographystyle{aa}
\bibliography{biblio}

\begin{thebibliography}{44}
\expandafter\ifx\csname natexlab\endcsname\relax\def\natexlab#1{#1}\fi

\bibitem[{{Andersen}(1991)}]{Andersen1991}
{Andersen}, J. 1991, \aapr, 3, 91

\bibitem[{{Asplund} {et~al.}(2009){Asplund}, {Grevesse}, {Sauval}, \&
  {Scott}}]{AGSS09}
{Asplund}, M., {Grevesse}, N., {Sauval}, A.~J., \& {Scott}, P. 2009, \araa, 47,
  481

\bibitem[{{Borucki} {et~al.}(2010){Borucki}, {Koch}, {Basri}, {Batalha},
  {Brown}, {Caldwell}, {Caldwell}, {Christensen-Dalsgaard}, {Cochran},
  {DeVore}, {Dunham}, {Dupree}, {Gautier}, {Geary}, {Gilliland}, {Gould},
  {Howell}, {Jenkins}, {Kondo}, {Latham}, {Marcy}, {Meibom}, {Kjeldsen},
  {Lissauer}, {Monet}, {Morrison}, {Sasselov}, {Tarter}, {Boss}, {Brownlee},
  {Owen}, {Buzasi}, {Charbonneau}, {Doyle}, {Fortney}, {Ford}, {Holman},
  {Seager}, {Steffen}, {Welsh}, {Rowe}, {Anderson}, {Buchhave}, {Ciardi},
  {Walkowicz}, {Sherry}, {Horch}, {Isaacson}, {Everett}, {Fischer}, {Torres},
  {Johnson}, {Endl}, {MacQueen}, {Bryson}, {Dotson}, {Haas}, {Kolodziejczak},
  {Van Cleve}, {Chandrasekaran}, {Twicken}, {Quintana}, {Clarke}, {Allen},
  {Li}, {Wu}, {Tenenbaum}, {Verner}, {Bruhweiler}, {Barnes}, \&
  {Prsa}}]{Borucki2010}
{Borucki}, W.~J., {Koch}, D., {Basri}, G., {et~al.} 2010, Science, 327, 977

\bibitem[{{Bressan} {et~al.}(2012){Bressan}, {Marigo}, {Girardi}, {Salasnich},
  {Dal Cero}, {Rubele}, \& {Nanni}}]{Bressan2012}
{Bressan}, A., {Marigo}, P., {Girardi}, L., {et~al.} 2012, \mnras, 427, 127

\bibitem[{{Brott} \& {Hauschildt}(2005)}]{brott05}
{Brott}, I. \& {Hauschildt}, P.~H. 2005, in ESA Special Publication, Vol. 576,
  The Three-Dimensional Universe with Gaia, ed. {C.~Turon, K.~S.~O'Flaherty, \&
  M.~A.~C.~Perryman}, 565--+

\bibitem[{{Caputo} {et~al.}(1989){Caputo}, {Chieffi}, {Tornambe}, {Castellani},
  \& {Pulone}}]{caputo1989}
{Caputo}, F., {Chieffi}, A., {Tornambe}, A., {Castellani}, V., \& {Pulone}, L.
  1989, \apj, 340, 241

\bibitem[{{Cassisi} {et~al.}(2001){Cassisi}, {Castellani}, {Degl'Innocenti},
  {Piotto}, \& {Salaris}}]{cassisi2001}
{Cassisi}, S., {Castellani}, V., {Degl'Innocenti}, S., {Piotto}, G., \&
  {Salaris}, M. 2001, \aap, 366, 578

\bibitem[{{Castellani} {et~al.}(1985){Castellani}, {Chieffi}, {Tornambe}, \&
  {Pulone}}]{castellani1985}
{Castellani}, V., {Chieffi}, A., {Tornambe}, A., \& {Pulone}, L. 1985, \apj,
  296, 204

\bibitem[{{Castellani} {et~al.}(1971){Castellani}, {Giannone}, \&
  {Renzini}}]{castellani1971}
{Castellani}, V., {Giannone}, P., \& {Renzini}, A. 1971, \apss, 10, 355

\bibitem[{{Castelli} \& {Kurucz}(2003)}]{castelli03}
{Castelli}, F. \& {Kurucz}, R.~L. 2003, in IAU Symposium, Vol. 210, Modelling
  of Stellar Atmospheres, ed. {N.~Piskunov, W.~W.~Weiss, \& D.~F.~Gray}, 20P--+

\bibitem[{{Claret}(2007)}]{Claret2007}
{Claret}, A. 2007, \aap, 475, 1019

\bibitem[{{Claret}(2011)}]{Claret2011}
{Claret}, A. 2011, \aap, 526, A157

\bibitem[{{Claret} \& {Gimenez}(1995)}]{Claret1995}
{Claret}, A. \& {Gimenez}, A. 1995, \aap, 296, 180

\bibitem[{{Claret} \& {Torres}(2016)}]{Claret2016}
{Claret}, A. \& {Torres}, G. 2016, \aap, 592, A15

\bibitem[{{Degl'Innocenti} {et~al.}(2008){Degl'Innocenti}, {Prada Moroni},
  {Marconi}, \& {Ruoppo}}]{scilla2008}
{Degl'Innocenti}, S., {Prada Moroni}, P.~G., {Marconi}, M., \& {Ruoppo}, A.
  2008, \apss, 316, 25

\bibitem[{{Deheuvels} {et~al.}(2016){Deheuvels}, {Brand{\~a}o}, {Silva
  Aguirre}, {Ballot}, {Michel}, {Cunha}, {Lebreton}, \&
  {Appourchaux}}]{Deheuvels2016}
{Deheuvels}, S., {Brand{\~a}o}, I., {Silva Aguirre}, V., {et~al.} 2016, \aap,
  589, A93

\bibitem[{{Dell'Omodarme} {et~al.}(2012){Dell'Omodarme}, {Valle},
  {Degl'Innocenti}, \& {Prada Moroni}}]{database2012}
{Dell'Omodarme}, M., {Valle}, G., {Degl'Innocenti}, S., \& {Prada Moroni},
  P.~G. 2012, A\&A, 540, A26

\bibitem[{{Gallenne} {et~al.}(2016){Gallenne}, {Pietrzyński}, {Graczyk},
  {Konorski}, {Kervella}, {Mérand}, {Gieren}, {Anderson}, \&
  {Villanova}}]{Gallenne2015}
{Gallenne}, A., {Pietrzyński}, G., {Graczyk}, D., {et~al.} 2016, \aap, 586,
  A35

\bibitem[{{Gennaro} {et~al.}(2010){Gennaro}, {Prada Moroni}, \&
  {Degl'Innocenti}}]{gennaro10}
{Gennaro}, M., {Prada Moroni}, P.~G., \& {Degl'Innocenti}, S. 2010, \aap, 518,
  A13+

\bibitem[{H{\"a}rdle \& Simar(2012)}]{simar}
H{\"a}rdle, W.~K. \& Simar, L. 2012, Applied Multivariate Statistical Analysis
  (Springer)

\bibitem[{{Hauschildt} {et~al.}(1999){Hauschildt}, {Allard}, \&
  {Baron}}]{hauschildt99}
{Hauschildt}, P.~H., {Allard}, F., \& {Baron}, E. 1999, \apj, 512, 377

\bibitem[{{Hauschildt} {et~al.}(2003){Hauschildt}, {Allard}, {Baron},
  {Aufdenberg}, \& {Schweitzer}}]{hauschildt03}
{Hauschildt}, P.~H., {Allard}, F., {Baron}, E., {Aufdenberg}, J., \&
  {Schweitzer}, A. 2003, in Astronomical Society of the Pacific Conference
  Series, Vol. 298, GAIA Spectroscopy: Science and Technology, ed. {U.~Munari},
  179--+

\bibitem[{{Jimenez} {et~al.}(2003){Jimenez}, {Flynn}, {MacDonald}, \&
  {Gibson}}]{jimenez03}
{Jimenez}, R., {Flynn}, C., {MacDonald}, J., \& {Gibson}, B.~K. 2003, Science,
  299, 1552

\bibitem[{{J{\o}rgensen} \& {Lindegren}(2005)}]{Jorgensen2005}
{J{\o}rgensen}, B.~R. \& {Lindegren}, L. 2005, \aap, 436, 127

\bibitem[{{Kirkby-Kent} {et~al.}(2016){Kirkby-Kent}, {Maxted}, {Serenelli},
  {Turner}, {Evans}, {Anderson}, {Hellier}, \& {West}}]{Kirkby-Kent2016}
{Kirkby-Kent}, J.~A., {Maxted}, P.~F.~L., {Serenelli}, A.~M., {et~al.} 2016,
  \aap, 591, A124

\bibitem[{{Noels} {et~al.}(2010){Noels}, {Montalban}, {Miglio}, {Godart}, \&
  {Ventura}}]{Noels2010}
{Noels}, A., {Montalban}, J., {Miglio}, A., {Godart}, M., \& {Ventura}, P.
  2010, \apss, 328, 227

\bibitem[{{Pagel} \& {Portinari}(1998)}]{pagel98}
{Pagel}, B.~E.~J. \& {Portinari}, L. 1998, \mnras, 298, 747

\bibitem[{{Paxton} {et~al.}(2013){Paxton}, {Cantiello}, {Arras}, {Bildsten},
  {Brown}, {Dotter}, {Mankovich}, {Montgomery}, {Stello}, {Timmes}, \&
  {Townsend}}]{MESA2013}
{Paxton}, B., {Cantiello}, M., {Arras}, P., {et~al.} 2013, \apjs, 208, 4

\bibitem[{{Peimbert} {et~al.}(2007{\natexlab{a}}){Peimbert}, {Luridiana}, \&
  {Peimbert}}]{peimbert07a}
{Peimbert}, M., {Luridiana}, V., \& {Peimbert}, A. 2007{\natexlab{a}}, \apj,
  666, 636

\bibitem[{{Peimbert} {et~al.}(2007{\natexlab{b}}){Peimbert}, {Luridiana},
  {Peimbert}, \& {Carigi}}]{peimbert07b}
{Peimbert}, M., {Luridiana}, V., {Peimbert}, A., \& {Carigi}, L.
  2007{\natexlab{b}}, in Astronomical Society of the Pacific Conference Series,
  Vol. 374, From Stars to Galaxies: Building the Pieces to Build Up the
  Universe, ed. {A.~Vallenari, R.~Tantalo, L.~Portinari, \& A.~Moretti}, 81--+

\bibitem[{{Pietrinferni} {et~al.}(2004){Pietrinferni}, {Cassisi}, {Salaris}, \&
  {Castelli}}]{teramo04}
{Pietrinferni}, A., {Cassisi}, S., {Salaris}, M., \& {Castelli}, F. 2004, \apj,
  612, 168

\bibitem[{{Stancliffe} {et~al.}(2015){Stancliffe}, {Fossati}, {Passy}, \&
  {Schneider}}]{Stancliffe2015}
{Stancliffe}, R.~J., {Fossati}, L., {Passy}, J.-C., \& {Schneider}, F.~R.~N.
  2015, \aap, 575, A117

\bibitem[{{Stancliffe} {et~al.}(2016){Stancliffe}, {Fossati}, {Passy}, \&
  {Schneider}}]{Stancliffe2016}
{Stancliffe}, R.~J., {Fossati}, L., {Passy}, J.-C., \& {Schneider}, F.~R.~N.
  2016, \aap, 586, A119

\bibitem[{{Tognelli} {et~al.}(2011){Tognelli}, {Prada Moroni}, \&
  {Degl'Innocenti}}]{Tognelli2011}
{Tognelli}, E., {Prada Moroni}, P.~G., \& {Degl'Innocenti}, S. 2011, \aap, 533,
  A109+

\bibitem[{{Torres} {et~al.}(2010){Torres}, {Andersen}, \&
  {Gim{\'e}nez}}]{Torres2010}
{Torres}, G., {Andersen}, J., \& {Gim{\'e}nez}, A. 2010, \aapr, 18, 67

\bibitem[{{Valle} {et~al.}(2014){Valle}, {Dell'Omodarme}, {Prada Moroni}, \&
  {Degl'Innocenti}}]{scepter1}
{Valle}, G., {Dell'Omodarme}, M., {Prada Moroni}, P.~G., \& {Degl'Innocenti},
  S. 2014, \aap, 561, A125

\bibitem[{{Valle} {et~al.}(2015{\natexlab{a}}){Valle}, {Dell'Omodarme}, {Prada
  Moroni}, \& {Degl'Innocenti}}]{binary}
{Valle}, G., {Dell'Omodarme}, M., {Prada Moroni}, P.~G., \& {Degl'Innocenti},
  S. 2015{\natexlab{a}}, \aap, 579, A59

\bibitem[{{Valle} {et~al.}(2015{\natexlab{b}}){Valle}, {Dell'Omodarme}, {Prada
  Moroni}, \& {Degl'Innocenti}}]{bulge}
{Valle}, G., {Dell'Omodarme}, M., {Prada Moroni}, P.~G., \& {Degl'Innocenti},
  S. 2015{\natexlab{b}}, \aap, 577, A72

\bibitem[{{Valle} {et~al.}(2015{\natexlab{c}}){Valle}, {Dell'Omodarme}, {Prada
  Moroni}, \& {Degl'Innocenti}}]{eta}
{Valle}, G., {Dell'Omodarme}, M., {Prada Moroni}, P.~G., \& {Degl'Innocenti},
  S. 2015{\natexlab{c}}, \aap, 575, A12

\bibitem[{{Valle} {et~al.}(2016{\natexlab{a}}){Valle}, {Dell'Omodarme}, {Prada
  Moroni}, \& {Degl'Innocenti}}]{testW}
{Valle}, G., {Dell'Omodarme}, M., {Prada Moroni}, P.~G., \& {Degl'Innocenti},
  S. 2016{\natexlab{a}}, \aap, 587, A31

\bibitem[{{Valle} {et~al.}(2016{\natexlab{b}}){Valle}, {Dell'Omodarme}, {Prada
  Moroni}, \& {Degl'Innocenti}}]{overshooting}
{Valle}, G., {Dell'Omodarme}, M., {Prada Moroni}, P.~G., \& {Degl'Innocenti},
  S. 2016{\natexlab{b}}, \aap, 587, A16

\bibitem[{{VandenBerg} {et~al.}(2006){VandenBerg}, {Bergbusch}, \&
  {Dowler}}]{VandenBerg2006}
{VandenBerg}, D.~A., {Bergbusch}, P.~A., \& {Dowler}, P.~D. 2006, \apjs, 162,
  375

\bibitem[{Venables \& Ripley(2002)}]{venables2002modern}
Venables, W. \& Ripley, B. 2002, Modern applied statistics with S, Statistics
  and computing (Springer)

\bibitem[{{Viallet} {et~al.}(2015){Viallet}, {Meakin}, {Prat}, \&
  {Arnett}}]{Viallet2015}
{Viallet}, M., {Meakin}, C., {Prat}, V., \& {Arnett}, D. 2015, \aap, 580, A61

\end{thebibliography}

\end{document}